\documentclass[11pt,a4paper]{article}

\usepackage[margin=2.5cm]{geometry}
\usepackage{amsmath,amssymb,bm}
\usepackage{graphicx,float,multirow,subcaption}
\usepackage[numbers]{natbib}
\usepackage{hyperref}

\title{The influence of quantum scarring on the onset of Bohmian chaos}
\author{A.C. Tzemos\thanks{Corresponding Author: atzemos@academyofathens.gr} and G. Contopoulos\\[0.5em]
\small Research Center for Astronomy and Applied Mathematics of the Academy of Athens,\\ Soranou Efessiou 4, GR-11527 Athens, Greece}
\date{}

\begin{document}
\maketitle

\begin{abstract}
In the present work, we investigate how the stable and unstable periodic orbits of a perturbed two–dimensional oscillator influence the Bohmian trajectories of the corresponding quantum system. In particular, we find that in  our systems  all Bohmian trajectories become chaotic in the long run and examine the  times  needed for them to approach  their ergodic pattern. We show that these times are systematically shorter for orbits close to unstable periodic orbits than for orbits close to stable periodic orbits of the same period. We demonstrate that this behaviour is connected to scarring effects associated with both unstable and stable periodic orbits. We also investigate how the Bohmian dynamics changes when a stable periodic orbit undergoes a bifurcation and becomes unstable.

\end{abstract}

\noindent\textbf{Keywords:} \texttt{Quantum Chaos, Hamiltonian Chaos, Bohmian Mechanics}


\section{Introduction}

In classical mechanics, chaos arises in nonlinear systems that exhibit extreme sensitivity to initial conditions.  
Trajectories that start arbitrarily close in phase space diverge exponentially with time, leading to a loss of predictability despite the deterministic nature of the equations of motion.  
Classical chaotic motion is characterized by positive Lyapunov exponents \cite{wiggins2003introduction}.

{
A similar perspective in the quantum regime is offered by Bohmian quantum mechanics, which provides a trajectory-based formulation equivalent to standard quantum theory \cite{Bohm,BohmII,holland1995quantum,sanz2012trajectory,sanz2013trajectory}. This equivalence is obtained when the initial distribution of Bohmian configurations satisfies the Born rule, meaning that the probability distribution of the configurations is given by the squared modulus of the wave function, $|\Psi|^2$. Due to the equivariance property of Bohmian dynamics, if the Born rule is satisfied initially, it remains satisfied at all subsequent times. Under this condition, Bohmian mechanics yields the same statistical predictions as standard quantum mechanics. However, the Born rule is not regarded as an independent axiom of Bohmian mechanics, and its origin and justification have been the subject of extensive research in the Bohmian literature. In particular, significant efforts have been devoted to understanding whether the Born distribution can emerge dynamically from more general initial distributions \cite{valentini1991signalI,valentini1991signalII,durr1992quantum,
durr2,valentini2005dynamical,efth2017,durr2021typicality,drezet2021justifying}.}

{In this framework, the wave function $\Psi = R e^{iS/\hbar}$ defines a velocity field through the relation}
\begin{equation}
m\dot{\mathbf{r}} = \hbar Im(\nabla \Psi / \Psi),
\end{equation}
{allowing one to define deterministic trajectories in configuration space. The motion of these trajectories is governed by the sum of the classical potential and an additional quantum potential derived from $R$ \cite{holland1995quantum}. Because the dynamics of these trajectories can be analysed in the classical way, the presence of chaos can be quantified using the same classical tools, such as sensitivity to initial conditions and Lyapunov exponents. This makes the Bohmian formulation particularly suitable for exploring how quantum systems inherit, enhance, or suppress classical chaotic behaviour.}

The first study on the connection between classical periodic orbits and the corresponding behaviour of the quantum wavefunction was made by Gutzwiller \cite{gutzwiller1971periodic, gutzwiller2013chaos}.
Later, Heller \cite{heller1984bound} considered unstable periodic orbits in the stadium model and demonstrated that when a wavepacket repeatedly revisits such an orbit, constructive interference builds up along it, leaving a visible trace in certain eigenfunctions, a phenomenon he termed ``quantum scarring''.

In fact, quantum scars provide one of the most striking examples of the interplay between quantum dynamics and classical instability.
In a completely chaotic classical system, one would expect the eigenstates of the Schrödinger equation to spread uniformly over the available phase space.
However, some eigenstates show a clear enhancement of probability density along the paths of unstable periodic orbits of the classical dynamics.
This type of  scarring is specific to unstable orbits and represents a partial quantum memory of classical instability \cite{eckhardt1988quantum,bogomolny1988smoothed, berry1989quantum, eckhardt1989quantum,bellomo1994state, kaplan1998linear,kaplan1998recent}. Scars of this type have been observed in a broad range of systems  \cite{sridhar1991experimental,wilkinson1996observation,ge2024direct}.
{However, similar localization of the quantum probability density can also occur around stable periodic orbits (see, for example, \cite{bellomo1994state, founargiotakis1989periodic}).}  


In the present work, we use the term ``scarring'' in a broader sense than the original Heller-type  scarring associated with unstable periodic orbits, referring to the enhancement or localization of the quantum probability density in the vicinity of classical periodic-orbit structures, including the more diffuse density enhancement within the stability islands surrounding stable periodic orbits. This terminology does not imply that the two cases represent identical localization mechanisms; we distinguish between sharp localization along unstable orbits and the softer density enhancement associated with stable orbits. In particular, we examine how the dynamical properties associated with stable and unstable periodic orbits affect the rate at which the Bohmian trajectories explore their corresponding phase-space structures. To the best of our knowledge, this comparison has not been considered previously and constitutes the main focus of the present paper.

In the present paper we work with a 2D harmonic oscillator perturbed by $\epsilon xy^2$ for various values of the perturbation parameter and for two  irrational values of the ratio of the unperturbed frequencies $\omega_y/\omega_x$.  Here the unperturbed dynamics are  integrable, but the presence  of the perturbation breaks integrability and induces stable and unstable periodic orbits that were absent in the original system. Thus we have both stable and unstable periodic orbits that can give rise to perturbation induced scarring phenomena as  found in \cite{founargiotakis1989periodic} and \cite{keski2019quantum}. In these papers (as well as in \cite{luukko2016strong,keski2017controllable}) the ratio of the unperturbed frequencies was a rational number. However, similar results are found if the ratio is an irrational number.

Within the framework of our ongoing investigation on  the connection between standard quantum and Bohmian quantum chaos \cite{contopoulos2020chaos,tzemos2024comparison}, we compute the Bohmian trajectories associated with wavepackets centred around unstable and stable periodic  orbits.  
 
By systematically varying the degree of phase coherence of the initial state  we numerically determine the characteristic times required for the trajectories to reach ergodic behaviour. Thus we  demonstrate how the presence of quantum scarring differentially affects the Bohmian dynamics. Finally, we study the transition of a classical periodic orbit from stability to instability as the nonlinearity parameter varies, and examine the corresponding changes in the patterns of the Bohmian trajectories across this transition.

{
The structure of the paper is the following: In Section 2  we present the model in  the cases  $\omega_x=1, \omega_y=\sqrt{3}$ and $\omega_x=1,\omega_y=\sqrt{2}$ both  from a classical and a Bohmian perspective.
In  Section 3 we construct wavepackets around the classical stable and unstable periodic orbits and  compare  the time needed for the development of chaos in Bohmian trajectories in each case. Furthermore we show what happens in the case of a bifurcation of a classical stable periodic orbit to an unstable one of the same period. 
Then we consider briefly two  rational  cases $\omega_x=\omega_y=1$ and $\omega_x=2,\omega_y=1$  while in Section 4 we extend our calculation in a case with a much higher energy. Finally in Section 5 we draw our conclusions.  In Appendix A we discuss the role of the initial position of the center of the wavepacket in the scarring phenomena and the corresponding Bohmian trajectories. In  Appendix B we present some technical details on the accuracy of the calculations.}

\section{Our model with non-commensurable frequencies}
We consider a two-dimensional anisotropic harmonic oscillator with frequencies 
$\omega_x = 1$ and $\omega_y = \sqrt{3}$, perturbed by a nonlinear coupling term of the form 
$\epsilon\,x y^2$:

\begin{equation}
H = \frac{1}{2}(p_x^2 + p_y^2) 
  + \frac{1}{2}(\omega_x^2 x^2 + \omega_y^2 y^2) 
  + \epsilon\,x y^2=E,\label{ham}
\end{equation}
where $E$ is the total energy.

\begin{figure}[H]
\centering
\subcaptionbox{}{\includegraphics[scale=0.2]{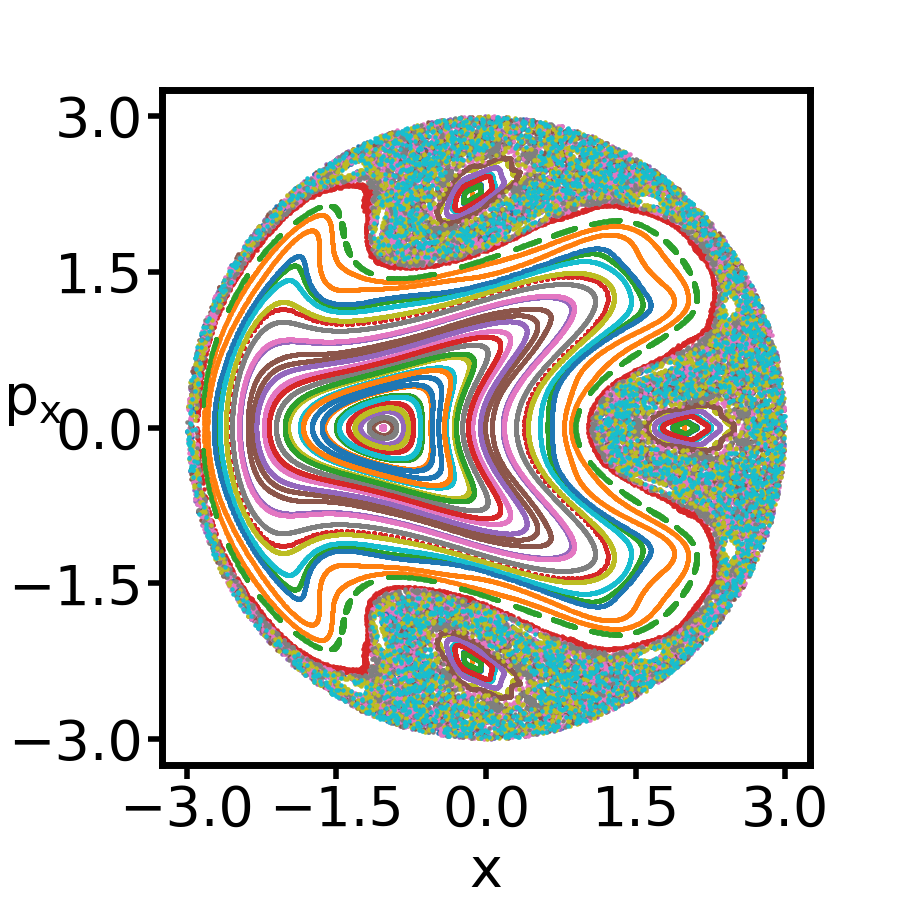}}
\subcaptionbox{}{\includegraphics[scale=0.2]{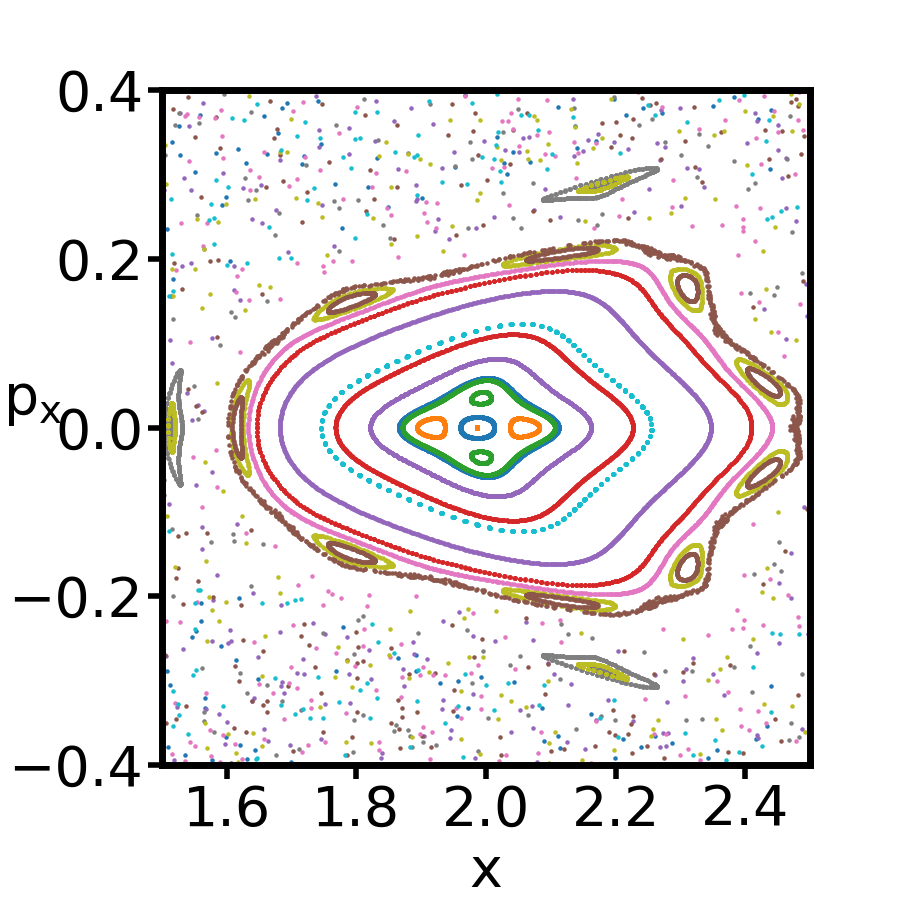}}
\subcaptionbox{}{\includegraphics[scale=0.2]{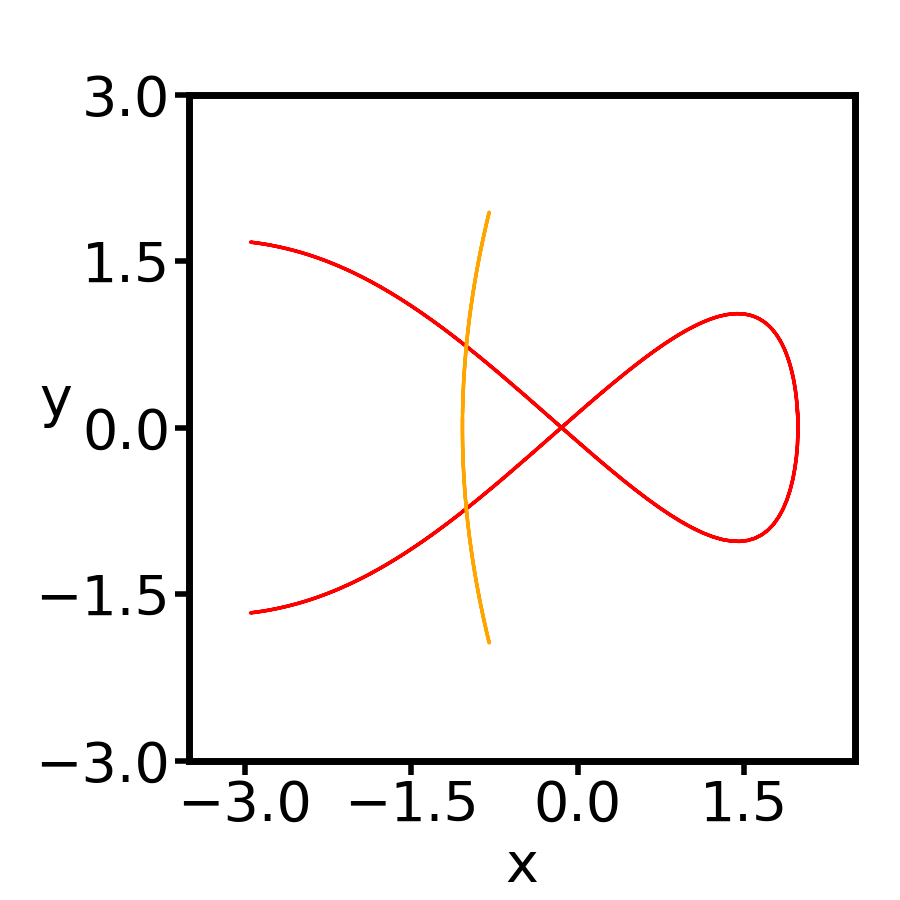}}
\subcaptionbox{}{\includegraphics[scale=0.2]{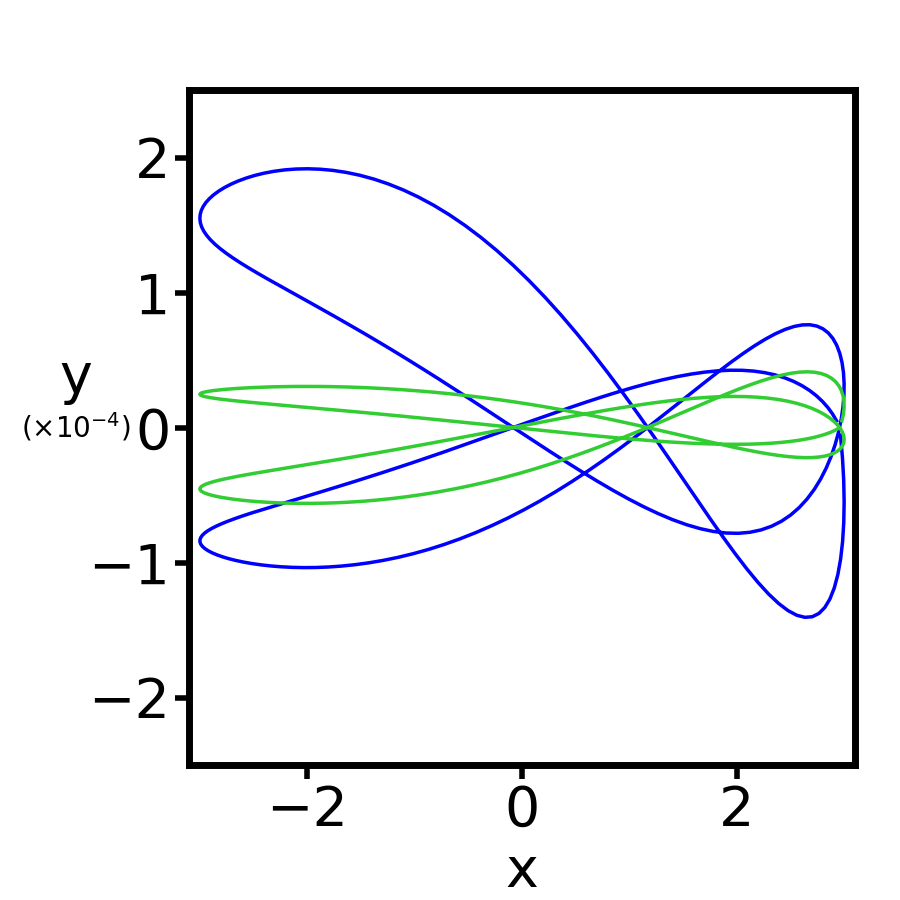}}
\caption{(a) Poincar\'{e} surface of section for $\epsilon=0.49$ and $E=4.4967$, (b) zoom in the right stability island whose center is a stable periodic orbit of period 3. (c) The stable period-1 (orange) and period-3 (red) periodic orbits. (d) The unstable period-3 periodic orbits with $p_x=-0.5012345$ (blue) and $p_x=0.5012345$ (green). We note that these orbits are very close to the axis $y=0$. }\label{fig1}
\end{figure}

Before examining the quantum eigenstates, we present the classical Poincaré surface of section $(x,p_x)$   at $y=0$ with $ p_y>0$ of the corresponding classical system (Fig.~\ref{fig1}(a)).

We consider first the case with $\epsilon=0.49$ and $E=4.4967$.
In Fig.~\ref{fig1}(a) we observe a large island of stability surrounding a period-1 stable periodic orbit  ($x=-1.036681948151
,p_x=0$), 
as well as three prominent secondary islands associated with a stable periodic orbit of period three. 
Inside the chaotic sea, additional smaller islands appear, corresponding to higher-order resonances (Fig.~\ref{fig1}(b)).  The whole Fig.~\ref{fig1}(a) is surrounded by a circle representing the periodic orbit $y=p_y=0$ or $x^2+p_x^2=2E$. The region around the right period-3 stability island is shown in Fig.~\ref{fig1}(b).  In Fig.~\ref{fig1}(c) we show two stable periodic orbits, one of period 1 (orange) and one of period 3 (red). There are also two period-3 unstable periodic orbits at $x=2.95672,p_x=\pm 0.50124$ which have a very small variation around the $y=0$ axis (Fig.~\ref{fig1}(d))\footnote{These two orbits differ in the value of $p_y$ at $y=0$. They represent two different sections of the same 4-d orbit  in opposite directions.}, as well as  higher order stable and unstable periodic orbits.

\begin{figure}[H]
\centering

\subcaptionbox{}{\includegraphics[scale=0.22]{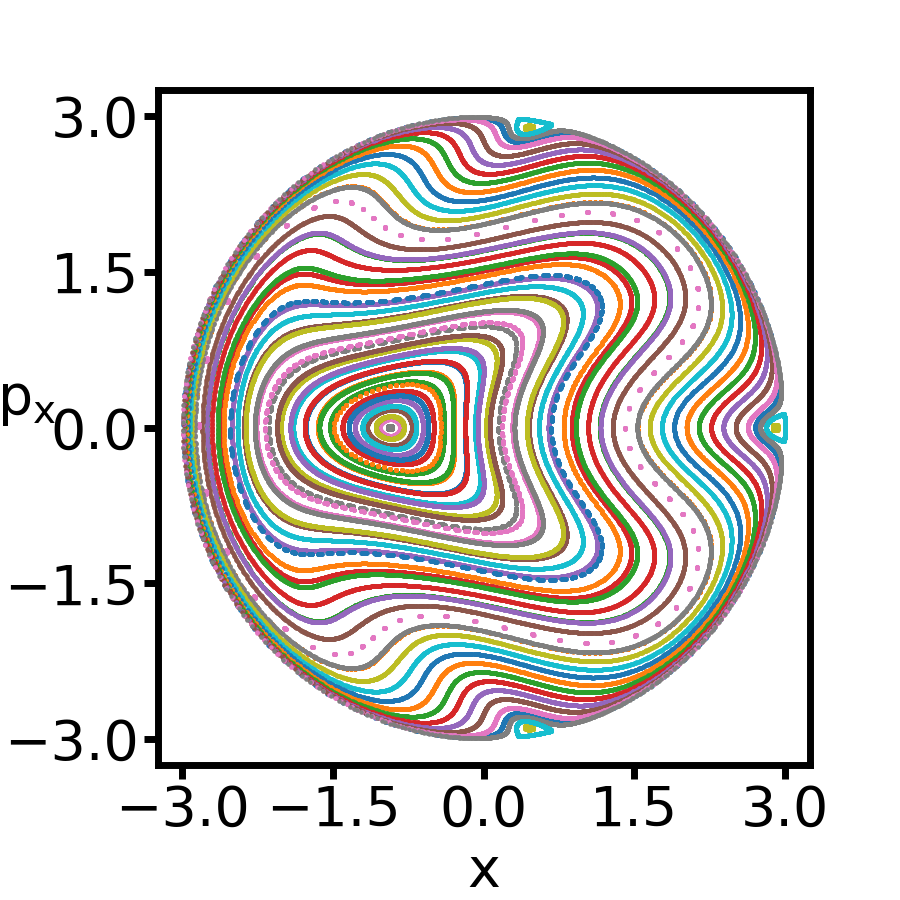}}
\subcaptionbox{}{\includegraphics[scale=0.22]{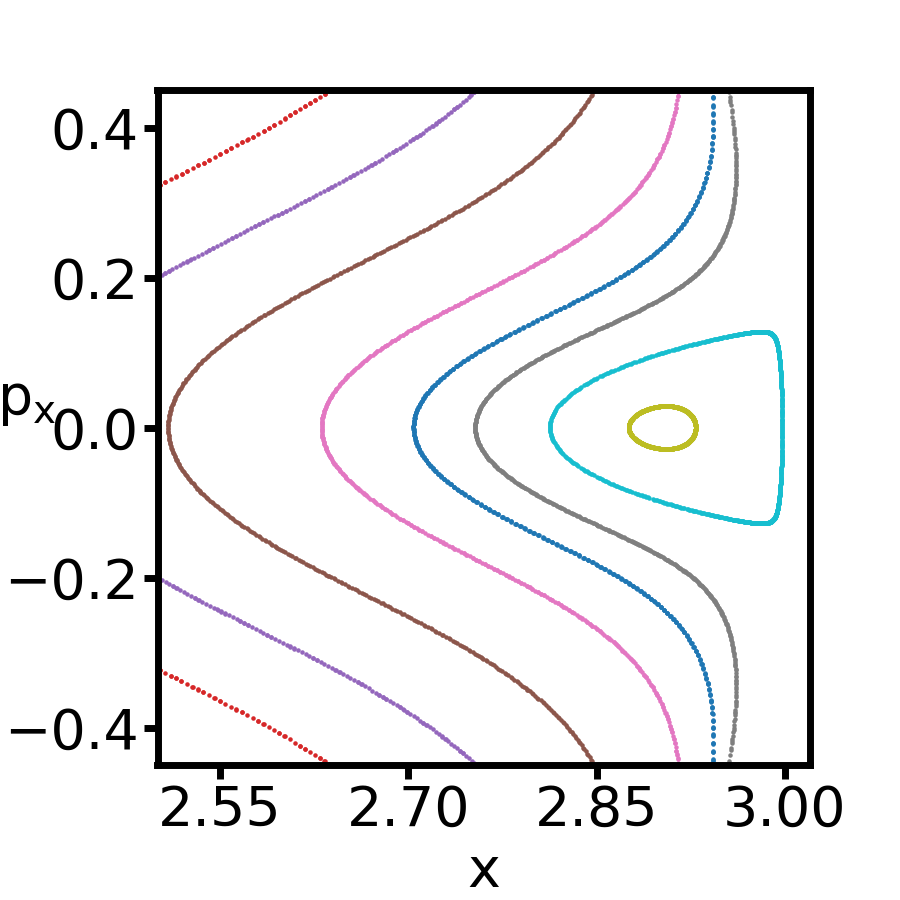}}
\subcaptionbox{}{\includegraphics[scale=0.22]{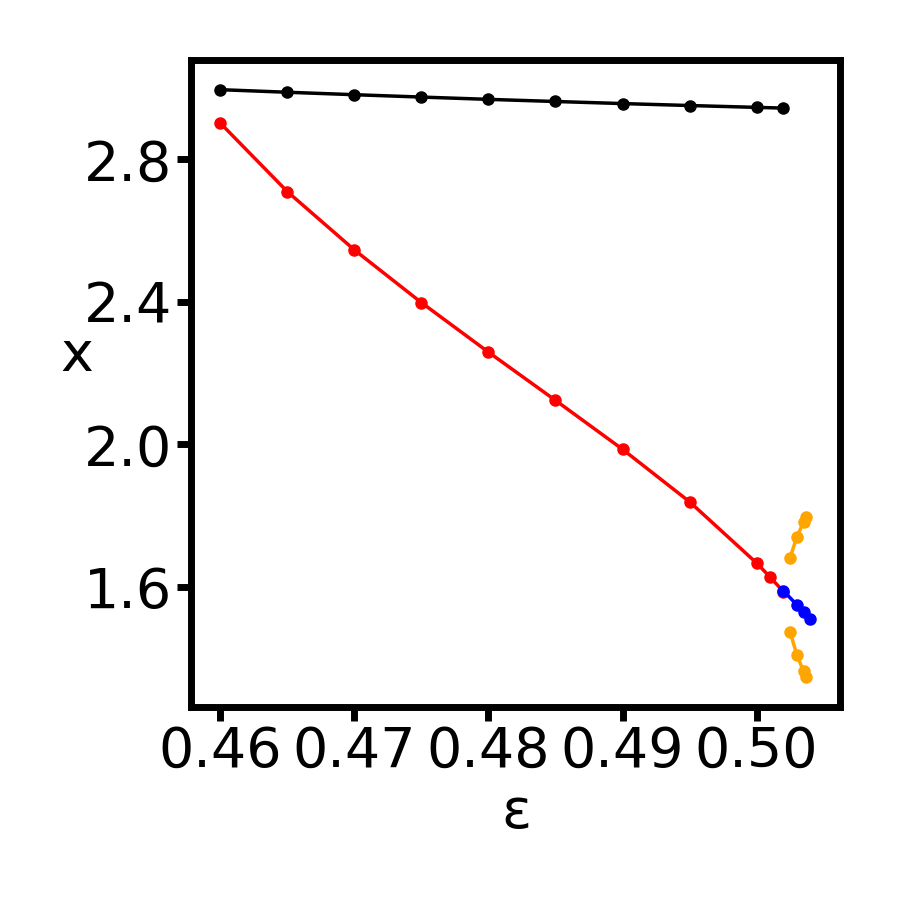}}
\subcaptionbox{}{\includegraphics[scale=0.22]{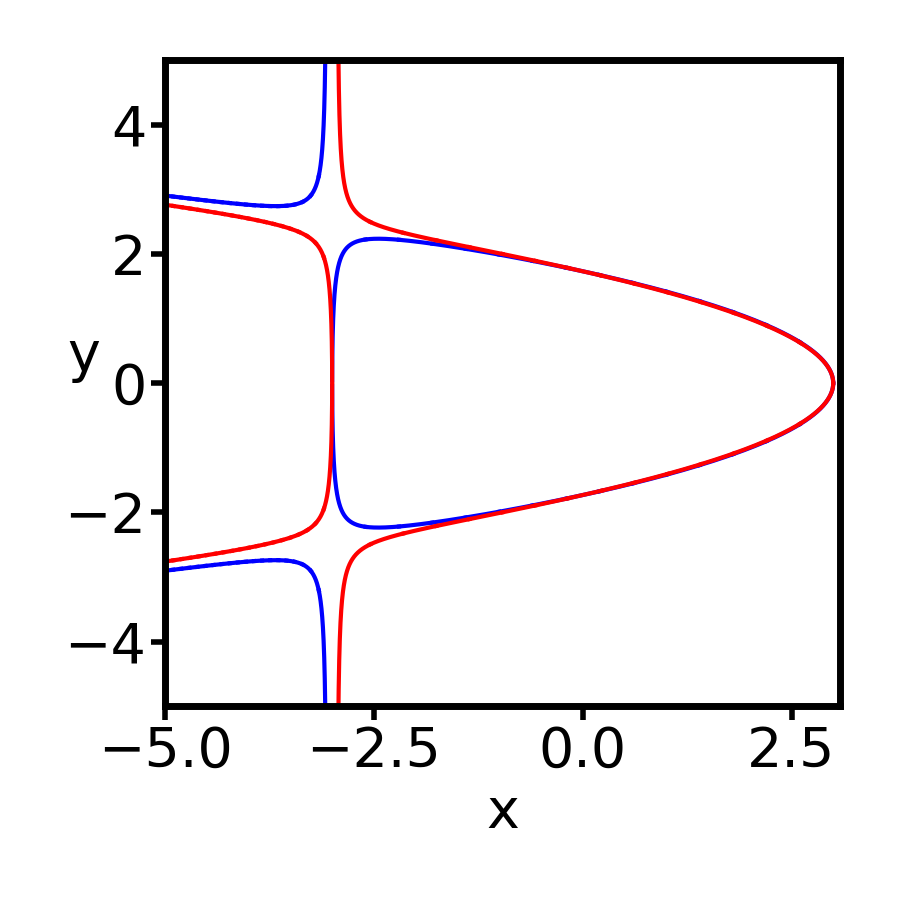}}
\subcaptionbox{}{\includegraphics[scale=0.22]{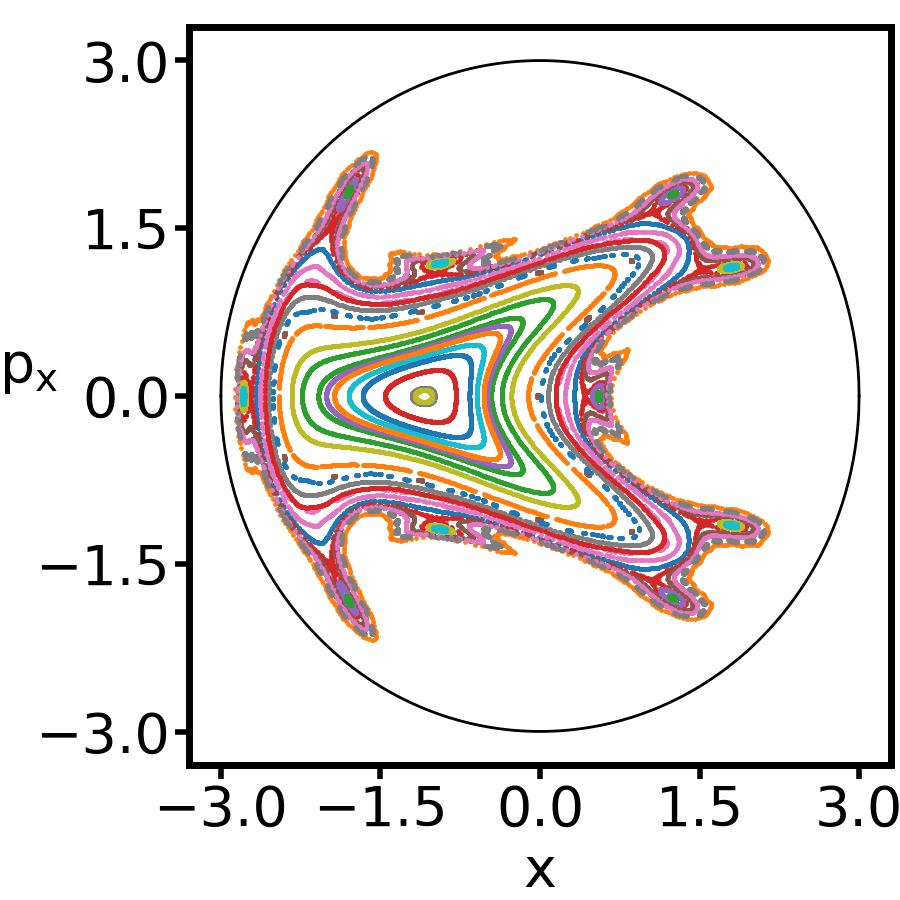}}

\caption{(a) The  surface of section for $\epsilon=0.46$ and energy $E=4.4967$. We see that the period-3 orbits are very close to the border of the section (b) A zoom of the right stability island. (c) The characteristic curve with respect to the perturbation $\epsilon$. We observe the bifurcation taking place at $\epsilon\simeq 0.5022$ (change from stable (red) to unstable (blue) and the generation of stable period-6 orbits (orange)). The black curve is the characteristic of the unstable period-3 orbit. (d) The  zero velocity curve (ZVC) for $\epsilon=0.49$ (blue, without escapes) and for $\epsilon=0.51$ (red, with escapes). (e) The surface of section for $\epsilon=0.505$ where we have escaping orbits (white regions). The surrounding black circle corresponds to the unstable periodic  orbit with $y=p_y=0$.}\label{fig2}
\end{figure}

\begin{figure}[H]
\centering
\subcaptionbox{}{\includegraphics[scale=0.2]{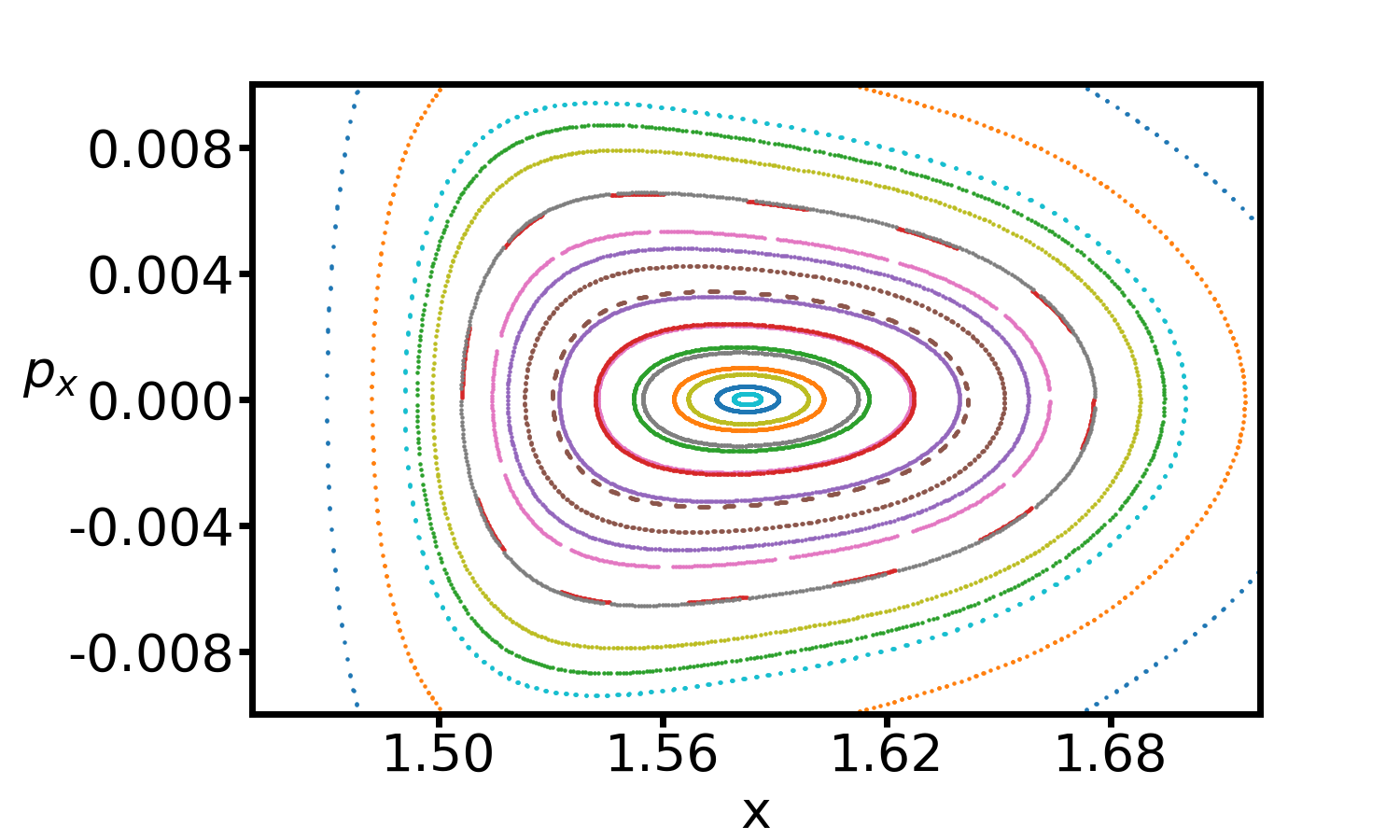}}
\subcaptionbox{}{\includegraphics[scale=0.2]{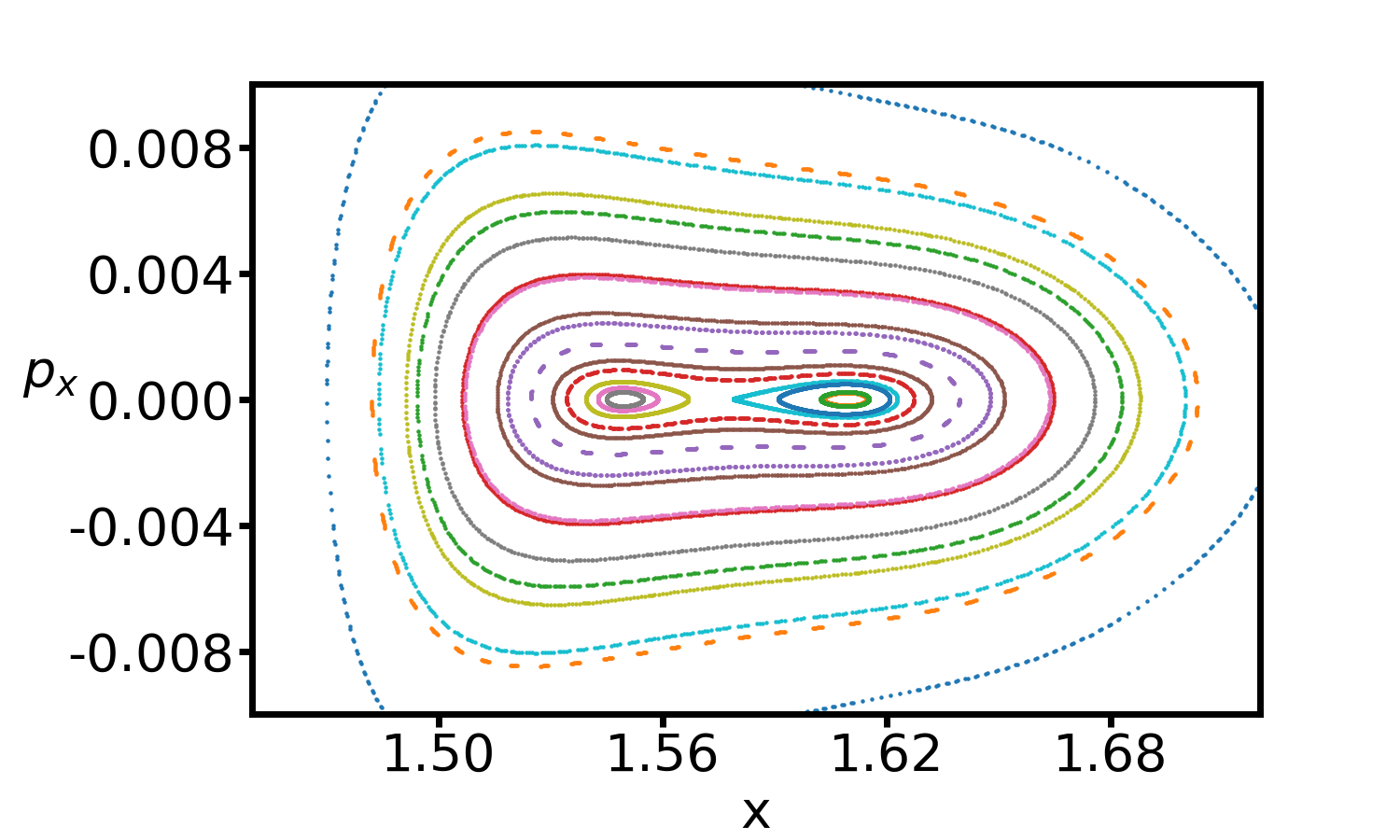}}
\caption{Two zoomed figures in the surface of section for (a) $\epsilon=0.5021$ and (b) $\epsilon=0.5022$ slightly before and after the bifurcation of the stable period-3 orbit to an unstable period-3 orbit and the simultaneous generation of 2 period-6 stable periodic orbits.}\label{fig3}
\end{figure}

The period-3 periodic orbits  have bifurcated from the periodic orbit $y=0$ (Eq.~\eqref{ham}) when $\epsilon$ was equal to $\epsilon=0.45.$ In fact, for   $\epsilon=0.46$, we see  the 3 points of the stable period-3 orbits  close to the boundary  (Figs.~\ref{fig2}(a),(b)) surrounded by islands of stability. Above and below every island there are two unstable periodic orbits of period 3, symmetric with respect to the $p_x$ axis very close to the boundary (Fig.~\ref{fig2}(b)). As $\epsilon$ increases the stable orbit  moves closer to the center. In Fig.~\ref{fig2}(c) we give the characteristics of the stable and of the two unstable orbits. The stable orbit (red curve) becomes unstable at about $\epsilon=0.502173$ generating two stable orbits of period 6 (Fig.~\ref{fig2}(c)). The two unstable periodic orbits (blue curve) (Fig.~\ref{fig2}(c)) remain very close to the boundary. For a little larger $\epsilon$ we start having escapes. In fact the curve of zero velocity ($p_x=p_y=0$)  is given by the equation
\begin{equation}
y^2=\frac{2E-\omega_x^2x^2}{\omega_y^2+2\epsilon x}
\end{equation} 
(we work with $\omega_x^2=1, \omega_y^2=3$) and for $x=-\sqrt{2E}$ it gives a zero denominator when \begin{equation}
\epsilon_{esc}=\frac{3}{2\sqrt{2E}}\simeq 0.5002.\end{equation} 
For slightly larger values of $\epsilon$ than $\epsilon_{esc}$ many orbits escape to infinity. The curve of zero velocity (red in Fig.~\ref{fig2}(d))  is open towards $y=\pm\infty$. For $\epsilon$ smaller than  $\epsilon_{esc}$ the curve of zero velocity forms a closed island (blue in Fig.~\ref{fig2}(d)) around the stable periodic orbit. The escaping orbits move  upwards or downwards. However, there are many non-escaping orbits around  the central period-1 orbit (Fig.~\ref{fig2}(e)). Near the transition from stable to unstable period-3 orbit the surface of section changes from Fig.~\ref{fig3}(a) (for $\epsilon=0.5021$) to Fig.~\ref{fig3}(b) ($\epsilon=0.5022)$. 
Beyond the first bifurcation at $\epsilon_1=0.502173$ we have a second bifurcation (producing 12 islands of stability) at $\epsilon_2=0.50412$. The interval between these two bifurcations is $\Delta\epsilon_1=0.001947$. Beyond the second bifurcation there is an infinity of period doubling bifurcations at intervals decreasing by successive powers of the universal ratio $\delta=8.7211$ \cite{contopoulos2002order} producing finally an infinity of unstable orbits beyond 
\begin{equation}
\epsilon_{\infty}=\epsilon_1+\frac{\Delta\epsilon_1}{1-1/\delta}=0.50437.
\end{equation}
As a consequence in Fig.~\ref{fig2}(e) for $\epsilon=0.505$ there is no island of stability left in the region around $x=1.5$. We note that the values of $\epsilon$ at the various bifurcations are slightly larger than the escape perturbation $\epsilon_{esc}$.

To study the quantum counterpart of Eq.~\eqref{ham}, we solve the stationary Schrödinger equation by diagonalizing the full perturbed Hamiltonian written 
in the basis of the unperturbed 2D harmonic oscillator eigenstates. 
The unperturbed eigenfunctions are given by
\begin{equation}
\Phi_{n_x,n_y}(x,y) =
\frac{1}{\sqrt{2^{n_x+n_y} n_x! n_y! \pi}} 
\, H_{n_x}(\sqrt{\omega_x}\,x)\,
H_{n_y}(\sqrt{\omega_y}\,y)\,
e^{-\frac{1}{2}(\omega_x x^2 + \omega_y y^2)},
\end{equation}
where $H_n$ denotes the Hermite polynomial of order $n$. 
Each perturbed eigenstate of the full Hamiltonian is expressed as a linear combination of these unperturbed basis functions
\begin{equation}
\tilde{\Phi}_k(x,y) = \sum_{n_x,n_y} C^{(k)}_{n_x,n_y}\,
\Phi_{n_x,n_y}(x,y),
\end{equation}
where the coefficients $C^{(k)}_{n_x,n_y}$ are obtained from the diagonalization of the Hamiltonian matrix in this basis.

An arbitrary initial wavefunction $\Psi_0(x,y)$ can then be expanded in the perturbed eigenbasis as
\begin{equation}
\Psi_0(x,y) = \sum_k a_k \tilde{\Phi}_k(x,y),
\end{equation}
and by substituting the expression of $\tilde{\Phi}_k$ in terms of the unperturbed functions, 
we obtain
\begin{equation}
\Psi_0(x,y) = \sum_k \sum_{n_x,n_y} 
a_k\, C^{(k)}_{n_x,n_y}\, \Phi_{n_x,n_y}(x,y).
\end{equation}
Here, the coefficients $a_k = \langle \tilde{\Phi}_k | \Psi_0 \rangle$ determine the projection of the initial wavepacket 
onto the perturbed eigenstates, while the $C^{(k)}_{n_x,n_y}$ coefficients encode the mixing between unperturbed states 
caused by the nonlinear perturbation.

In principle, the summations over $n_x$ and $n_y$ in the above expressions extend to infinity, as the Hilbert space of the unperturbed two-dimensional harmonic oscillator is infinite-dimensional. In practice, however, we truncate the basis  by imposing the condition $n_x + n_y \leq N$. In the present calculations, we take $N = 10$, which corresponds to a total of 66 unperturbed basis states. This truncation has been carefully tested and shown to provide excellent accuracy: both the total probability normalization and the expectation value of the energy of an initial coherent wavepacket centered near the periodic orbit are preserved to within less than one percent over long-time evolution. Similar convergence is also observed when constructing arbitrary superpositions of the perturbed eigenstates.

Furthermore, increasing the truncation (for instance, up to $n_x + n_y = 28$, corresponding to 435 unperturbed basis states) does not modify the spectral structure in the energy region of interest. In both truncations, only three perturbed eigenstates are found in the vicinity of the classical energy of the unstable period-three orbit ($E_{\text{orbit}} \approx 4.5$).   In fact, we label the energies of the perturbed eigenstates  $\tilde{\Phi}_k$ in ascending order then  $\tilde{E}_6=4.288, \tilde{E}_7=4.518$ and $\tilde{E}_8=4.8525$. This similarity of the eigenvalue pattern confirms that the scarred multiplet is fully convergent and that the truncation does not affect the physical features associated with the quantum scarring phenomenon.

\begin{figure}[H]
\centering
\subcaptionbox{}{\includegraphics[scale=0.173]{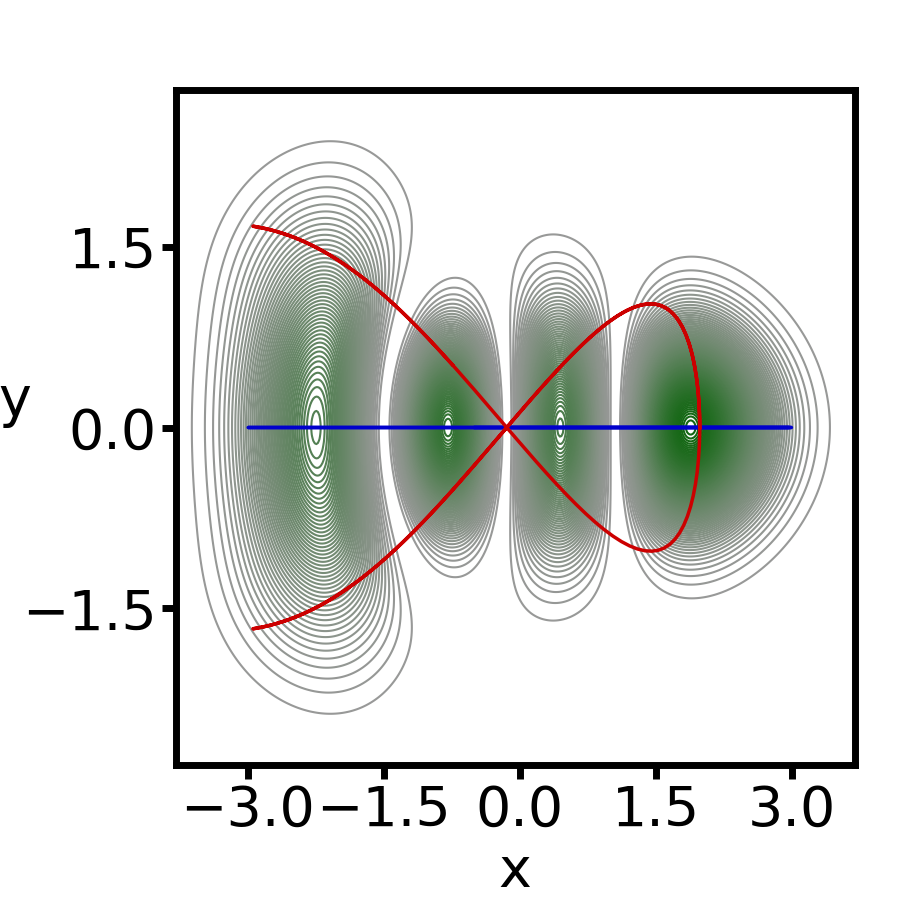}}
\subcaptionbox{}{\includegraphics[scale=0.173]{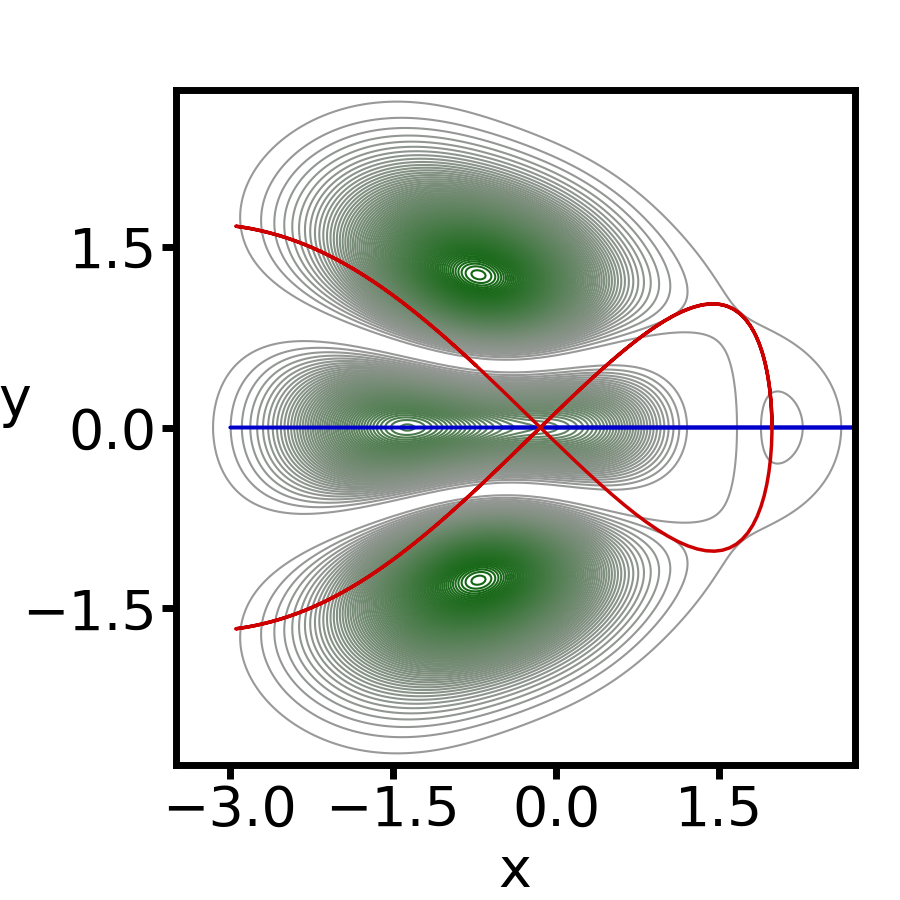}}
\subcaptionbox{}{\includegraphics[scale=0.173]{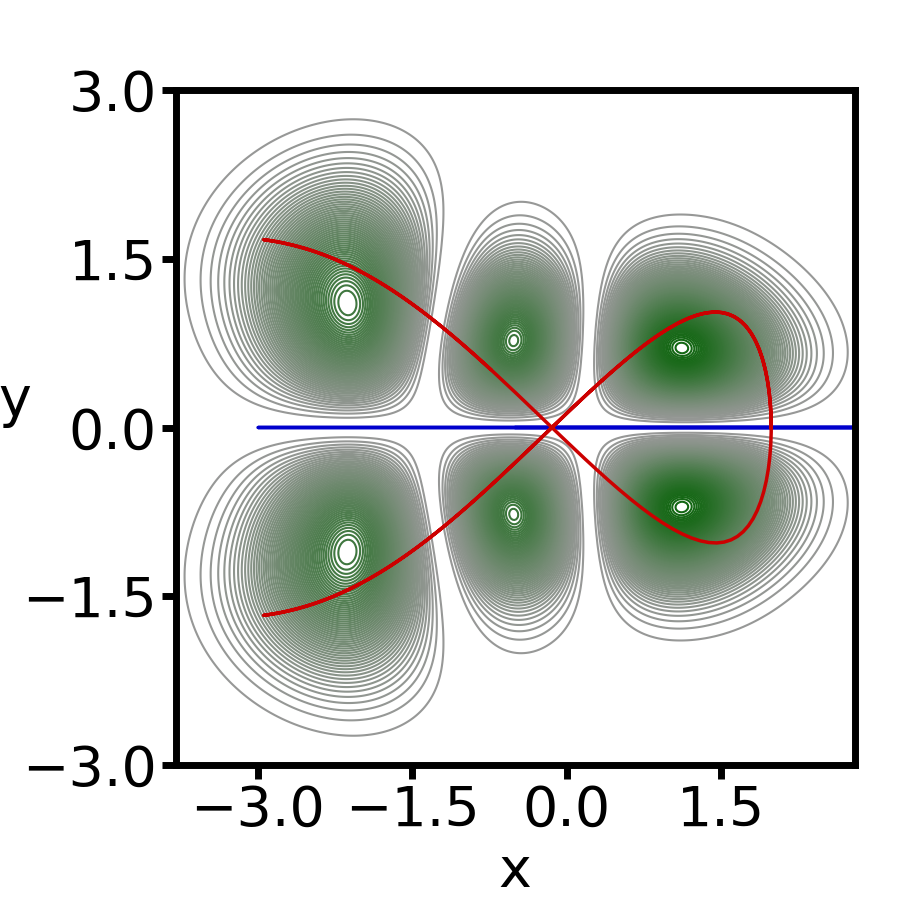}}
\includegraphics[scale=0.22]{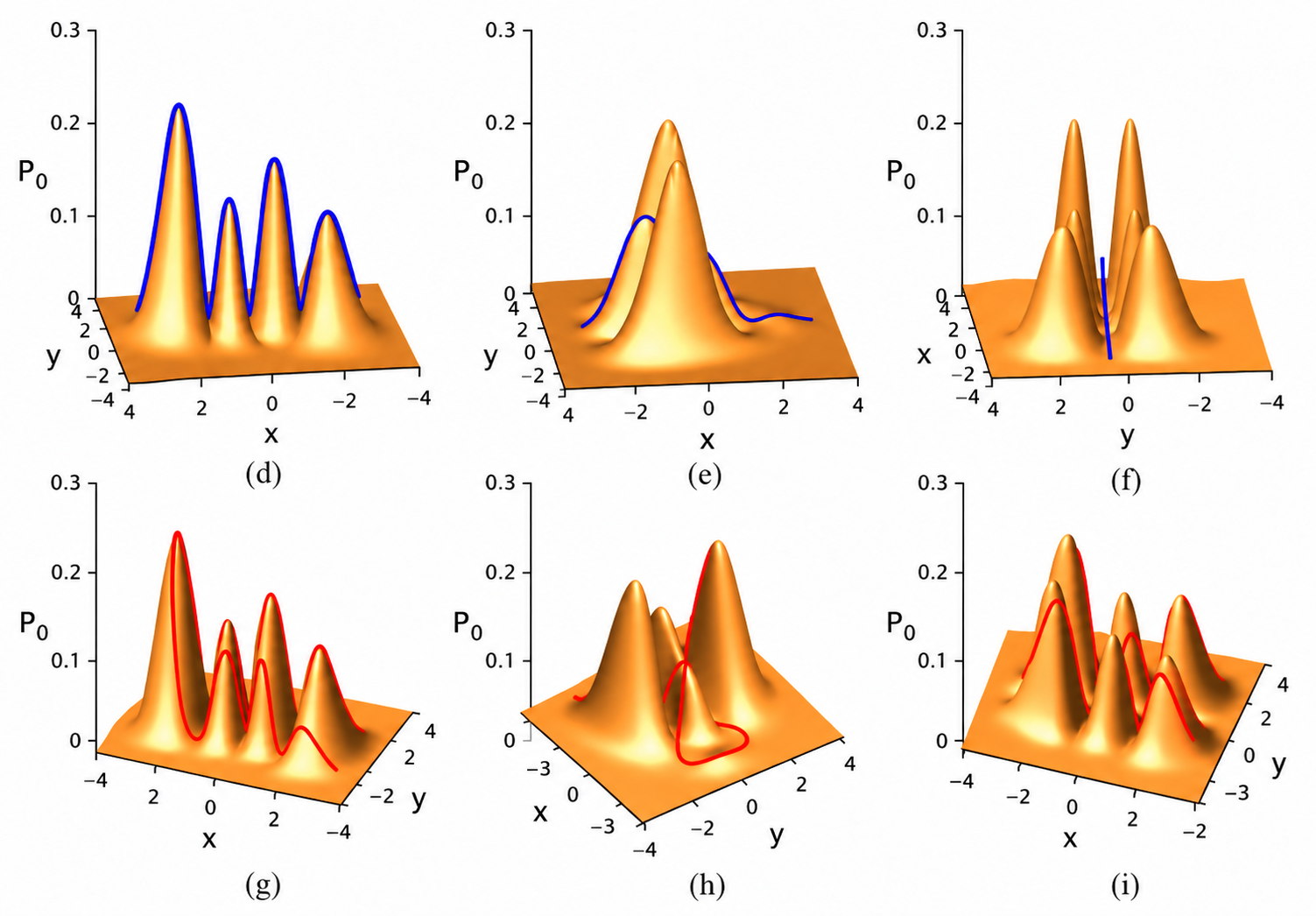}
\caption{Contour plots of the  probability densities of the three eigenstates of the perturbed system with the closest energies to that of the Poincar\'{e} surface of section: (a) $E=4.288$, (b) $E=4.518$ and (c) $E=4.852$. Superimposed we see the stable (red) and the unstable (blue) periodic orbits of period 3. In (d)-(i) we show the corresponding 3D analogues of (a)-(c). }\label{fig4}
\end{figure}

In Figs.~\ref{fig4}(a),(b),(c), we present the 2D density plots of $|\tilde{\Phi}_k(x,y)|^2$ for $k=6,7,8$. We observe that they exhibit symmetry with respect to the $y=0$ axis, which arises from the invariance of the perturbation term $\epsilon xy^2$ under the transformation $y \to -y$.

The unstable orbit of period 3 (blue curve) has initial position on the surface of section  $x(0)=2.9567, y(0)=0, p_x(0)=-0.5012, p_y(0)=0.0001$. Its initial condition  is very close to the border of the section of Fig.~\ref{fig1}(a). The orbit is very thin in the y direction and very close to the $y=0$ axis. Thus we expect from the density plots that only $\tilde{\Phi}_6$ and $\tilde{\Phi}_7$ will feel its presence. This is shown clearly in Figs.~\ref{fig4}(d),(e) where we have the 3D probability density plots corresponding to the 2D plots along with the unstable orbit projected on them. It is clear that the unstable periodic orbit passes very close to the  tops of all blobs of $|\tilde{\Phi}_6|^2$ which are all placed at the $y=0$ axis. This curve also passes very close to the tops of the two  central blobs of $|\tilde{\Phi}_7|^2$, which are along the $y=0$ axis. However, the density  $|\tilde{\Phi}_8|^2$ is not supported by the unstable periodic orbit since this orbit is located at the bottom of the valley between the various blobs.

A similar inspection can be made for the stable period–3 orbit, shown in Figs.~\ref{fig4}(g),(h),(i). In this case we observe that the scarring represented by the blobs is supported by the stable periodic orbits, but not with the tops of the blobs very close to the periodic orbit as in the case of the unstable periodic orbit. The nearby eigenstates of the perturbed Hamiltonian contribute to the density $P=|\Psi|^2$ inside the associated stability island, yet this enhancement is spread over the whole region rather than concentrated along the periodic orbit. In $|\tilde{\Phi}_6|^2$ the stable orbit crosses the four blobs without aligning with their peaks; in $|\tilde{\Phi}_7|^2$ it passes through the two central and the upper–lower lobes without  approaching their tops. In $|\tilde{\Phi}_8|^2$ it passes through four out of the six blobs. Consequently, these features represent a softer  type of scarring generated by the stable orbit, in contrast to the sharp scarring produced by the unstable orbit.

It is worth emphasizing that, in the case of stable periodic orbits, the correspondence between the periodic orbit and the density enhancement is not expected to be as localized as for an unstable periodic orbit. The probability density generally contains several maxima and minima, and some of the maxima may occur away from the classical periodic orbit. The relevant feature is therefore not that every density maximum lies directly on the orbit, but rather that the periodic orbit traverses the enhanced-density structures associated with the scarring pattern. In this sense, the stable-orbit structures observed here should be regarded as a broader and softer form of scarring, characterized by an enhancement of probability density in the region surrounding the periodic orbit rather than by a sharp concentration directly along its path.

\section{Wavepacket  Bohmian Dynamics}

\subsection{Non-resonant cases}
Individual energy eigenstates are stationary and generally do not provide a localized quantum representation of a particle moving along a classical trajectory. To establish the closest possible connection with classical motion, we therefore employ Gaussian coherent wavepackets constructed as superpositions of perturbed energy eigenstates and centred on selected classical periodic orbits. Such a  wavepacket has the general form:
\begin{equation}\label{wp}
\Psi_0^{\mathrm{coh}}(x,y)
= \frac{1}{\sqrt{\pi \sigma_x \sigma_y}}
\exp\!\left[-\frac{(x - x_0)^2}{2\sigma_x^2}
- \frac{(y - y_0)^2}{2\sigma_y^2}
+ \frac{i}{\hbar}(p_{x_{0}}x + p_{y_{0}}y)\right],
\end{equation}
and it is localized around the classical periodic orbit, providing a semiclassical representation of a single particle around the position $(x_0, y_0)$ of the periodic orbit with the corresponding momenta $(p_{x_0}, p_{y_0})$. 
The parameters $\sigma_x$ and $\sigma_y$ determine the spatial widths of the Gaussian packet along the $x$ and $y$ directions, respectively, and fix the corresponding momentum spread through Heisenberg's uncertainty relation.  
\textbf{In fact, for $\sigma_x = \sqrt{\hbar/(2m\omega_x)}$ and $\sigma_y = \sqrt{\hbar/(2m\omega_y)}$  
the above wavepacket starts with a  minimum uncertainty at $t=0$ (it is a product of two coherent states along the $x$ and $y$
axes)}. However, the presence of the perturbation term
$\epsilon xy^2$ destroys  coherence as the system evolves in time.

Such a state naturally includes a superposition of nearby perturbed eigenstates, among which are those that exhibit scarring close to the energy of the triply unstable periodic orbit. 
In individual perturbed eigenstates, scarring is a purely static phenomenon that manifests itself as a localized enhancement of probability density along a periodic orbit. 
In such stationary states, the Bohmian trajectories are time-independent and form a static flow field that reflects the spatial profile of the scar. 
Scarring therefore represents a form of spatial localization but does not by itself produce the necessary conditions for the emergence of Bohmian chaos.

{To generate  Bohmian chaos, temporal interference among several stationary components is required \cite{parmenter1995deterministic,contopoulos2020chaos,
makowski2001simplest,abdul2026frequency}. 
For a superposition of stationary eigenstates with real coefficients, at least three eigenstates of the Hamiltonian are required for the emergence of Bohmian chaos. For a superposition of two eigenstates with real coefficients, the Bohmian velocity field contains only a single independent energy difference, and therefore lacks the multiple-frequency temporal structure necessary for the emergence of Bohmian chaos. When the coefficients are complex, however, the additional relative phase between the two components can provide the necessary time-dependent phase structure, allowing chaotic Bohmian dynamics to arise even from a two-state superposition.}

In the present system, the initial Gaussian coherent wavepackets are projected onto the eigenbasis of the perturbed Hamiltonian, yielding a time-dependent superposition of a large number  of perturbed eigenstates (66 in this case)  with distinct eigenenergies and, in general, complex expansion coefficients, thereby generating the  interference structure required for Bohmian chaos.\footnote{ We note that the three eigenstates displayed in Fig.~\ref{fig4} are selected to illustrate the stationary probability-density patterns near the energy of the classical periodic orbit; they do not constitute the full superposition used in the dynamical simulations.}

We first constructed a coherent wave packet centred on the   periodic orbit in configuration space by setting the parameters $(x_0, y_0, p_{x0}, p_{y0})$ accordingly. 
An ensemble of 5000 Bohmian particles was then generated according to the Born rule distribution $|\Psi_0^{\mathrm{coh}}|^2$ and propagated under the perturbed Hamiltonian. 
In this configuration, the Bohmian trajectories exhibit a very rapid onset of chaos, with the particles quickly spreading throughout the accessible configuration space \footnote{The support of the wavefunction extends from $-\infty$ to $+\infty$, however beyond an area of $6\times 6$ the probability density is extremely small (smaller than $10^{-6}$).}

To quantify chaos in a dynamical system, the natural diagnostic is the Lyapunov characteristic number (LCN). For a trajectory $\mathbf{x}(t)$, the LCN is defined as
\begin{equation}
\lambda = \lim_{t \to \infty} \frac{1}{t}\,\ln \frac{\|\delta \mathbf{x}(t)\|}{\|\delta \mathbf{x}(0)\|},
\end{equation}
where $\delta \mathbf{x}(t)$ is an infinitesimal deviation vector evolved through the variational equations. A positive value of $\lambda$ indicates exponential sensitivity to initial conditions and therefore chaotic behaviour, while ordered trajectories are characterised by $\lambda=0$.

In Bohmian mechanics, however, the practical computation of LCNs for a large ensemble of trajectories is highly demanding. The stiffness of the Bohmian equations requires extremely long integration times  for the limit  $\lambda$ to converge reliably, making systematic LCN computations computationally prohibitive.

In our previous works \cite{contopoulos2020chaos,tzemos2020ergodicity}, we showed that chaotic Bohmian trajectories are ergodic, in the sense that they all fill the same effective support of the wavefunction and converge to essentially the same long-time point distribution, depicted by colorplots  that record how many times each trajectory visits each square bin of a grid covering the region where the probability density is non-negligible (in practice $|\Psi|^2>10^{-5}$). By computing the Frobenius distances \cite{strang1993introduction} between the underlying matrices of various colorplots we found that in all cases they tend to become practically zero. In contrast, ordered trajectories are non-ergodic, and cover only a small subset of the effective support.

\begin{figure}
\centering
\includegraphics[scale=0.21]{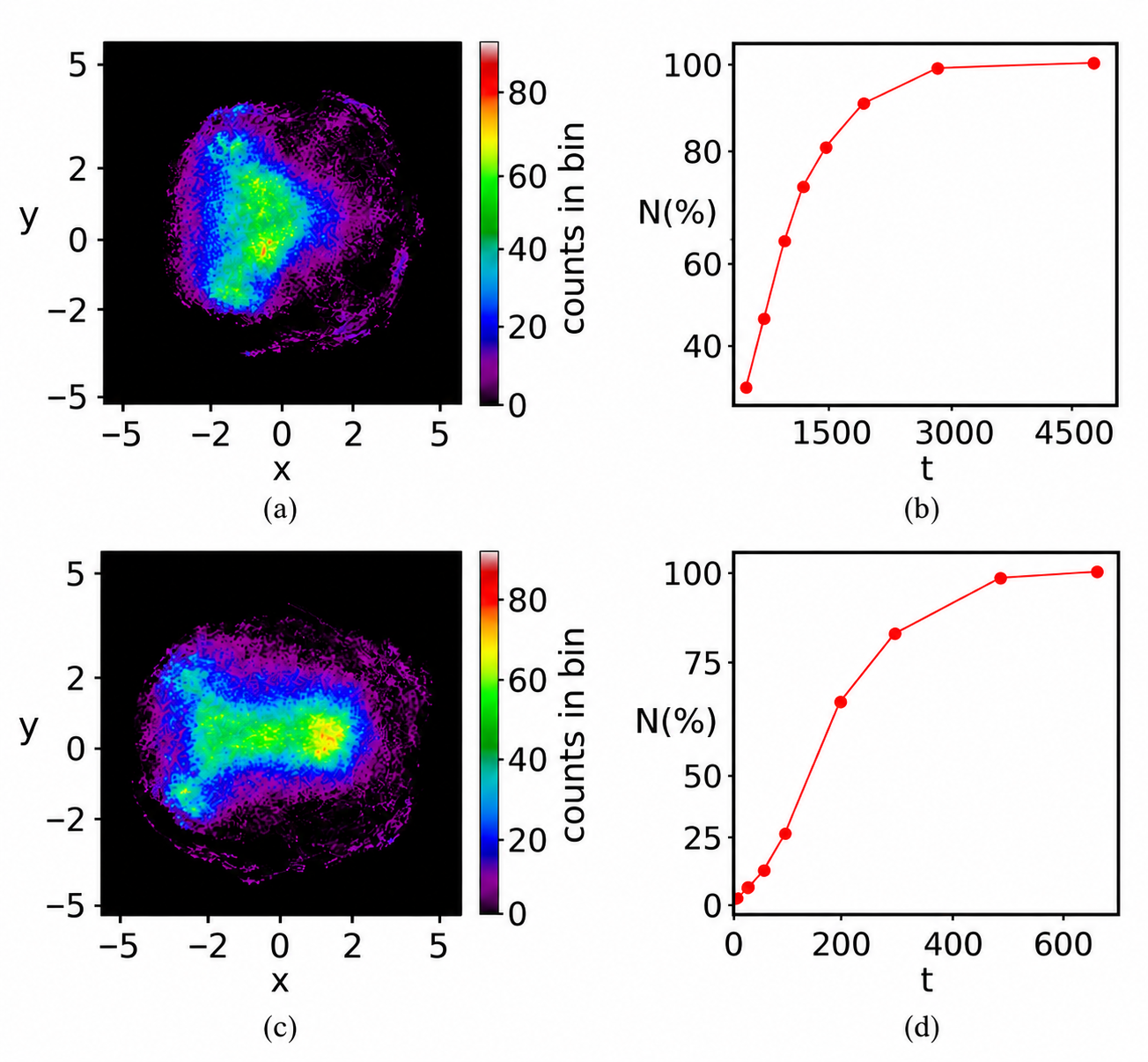}

\caption{(a) The colorplot of a single chaotic Bohmian trajectory for $\epsilon=0.49$ ($x(0)=3,y(0)=-0.1$) for $t=15000$ in the Born distribution of the wavepacket centred around the stable period-1 orbit. (b) The percentage $N$ of the Bohmian trajectories that have covered for the first time the area of the final ergodic pattern as a function of time for a wavepacket around the stable period-1 orbit. (c),(d) The same as (a),(b) in the case of the wavepacket around the stable period-3 orbit. We see that the times  required for the same proportion $N$ of trajectories around the  period-1 stable periodic orbit to reach ergodicity are much longer than those of the orbits around the period-3 stable periodic orbit.}\label{fig5}
\end{figure}

An example is given in Fig.~\ref{fig5}(a) where we show the color plot of a chaotic Bohmian trajectory inside the Born rule distribution of the wavepacket centred around the period-1 stable periodic orbit. The colorplot shows how many times the trajectory has passed each bin of a square grid with $\Delta x=\Delta y=0.05$ up to $t=15000$. This is essentially  common for all the chaotic trajectories of the corresponding wavepacket. If we set a filter and ask what is the percentage $N$ of Bohmian trajectories that have covered for the first time the boundary of an area of size 45 (major part of the colorplot in Fig.~\ref{fig5}(a)) as a function of time then we find the diagram of Fig.~\ref{fig5}(b). We see that $N$ goes to $100\%$ at about $t=3000$, i.e. all trajectories are chaotic and cover, for the first time, the major part of their long term ergodic pattern at about $3000$ time units. If instead we make a wavepacket around the period-3 stable periodic orbit, then we find the colorplot of Fig.~\ref{fig5}(c) and the corresponding times Fig.~\ref{fig5}(d). We see that the times required for any given proportion $N$ of orbits to reach approximately the area of the ergodic pattern is much larger for the orbits around the central (period-1) periodic orbit than for the orbits around the period-3 stable periodic orbit (Figs.~\ref{fig5}(b),(d)).

\begin{figure}[H]
\centering
\includegraphics[scale=0.17]{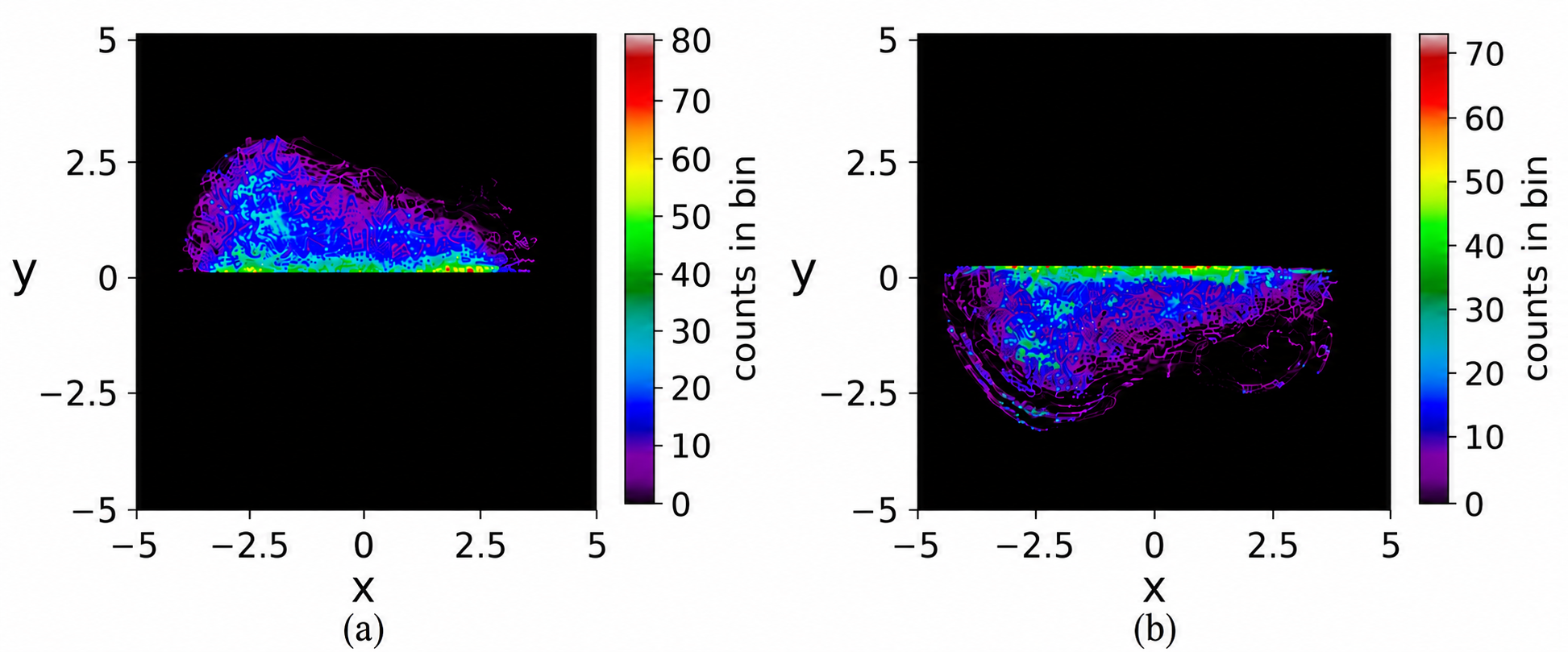}

\caption{  Two partially ergodic  Bohmian trajectories up to $t=5000$: (a) for $x(0)=1, y(0)=0.7$ and (b) for $x(0)=-0.9, y(0)=-0.6$. The trajectories never cross the $y=0$ axis.  }\label{fig6}
\end{figure}

Then we repeat the above process for a wavepacket centred at the period-3 unstable periodic orbit, where, unlike in the previous case, the Bohmian ensemble exhibits partial ergodicity \cite{tzemos2023unstable} with respect to the symmetry axis $y=0$.  Namely the orbits starting above this $y=0$ axis fill the upper  region ergodically  but do not go below the $y=0$ axis and the orbits below $y=0$ do not go above $y=0$. This behaviour is displayed in Figs.~\ref{fig6}(a),(b) which show the colorplots of two chaotic trajectories  above and below the axis $y=0$. In both situations all Bohmian trajectories have a common colorplot confined to the half-plane in which they originate. This behaviour stems from the fact that, in this special case, the y-component of the Bohmian field is zero on the $y$-axis $\forall t$.  We note that this behaviour occurs only for this particular periodic orbit.

\begin{figure}[H]
\centering

\subcaptionbox{}{\includegraphics[scale=0.21]{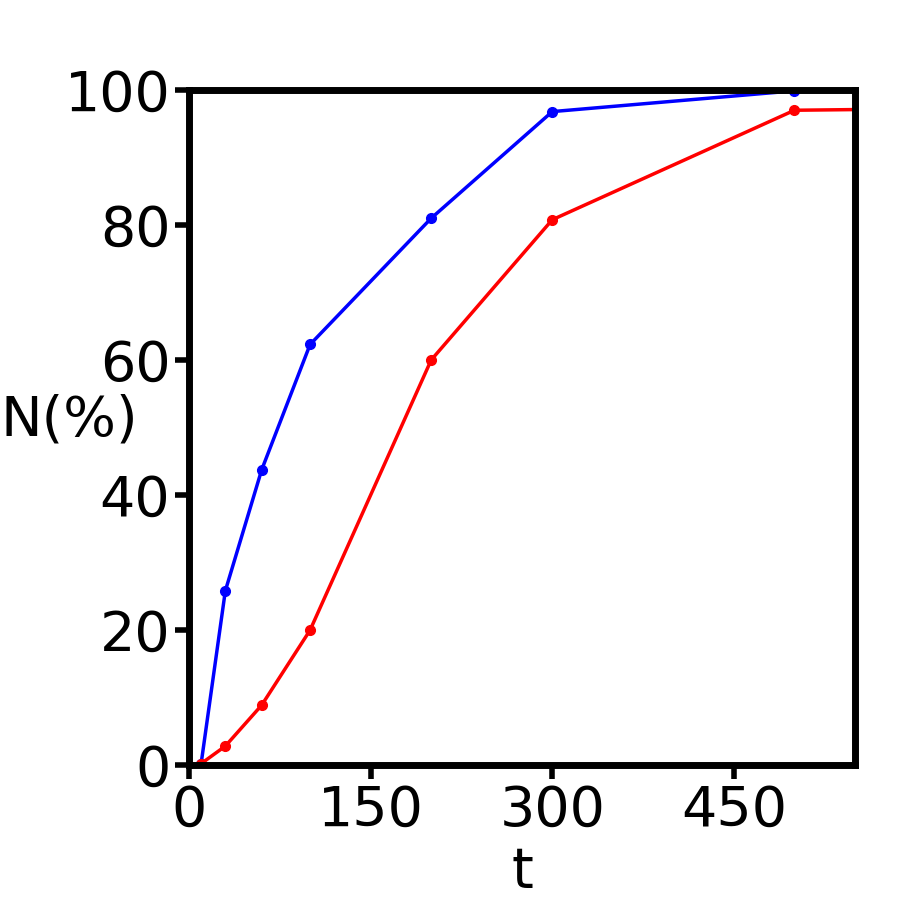}}
\subcaptionbox{}{\includegraphics[scale=0.21]{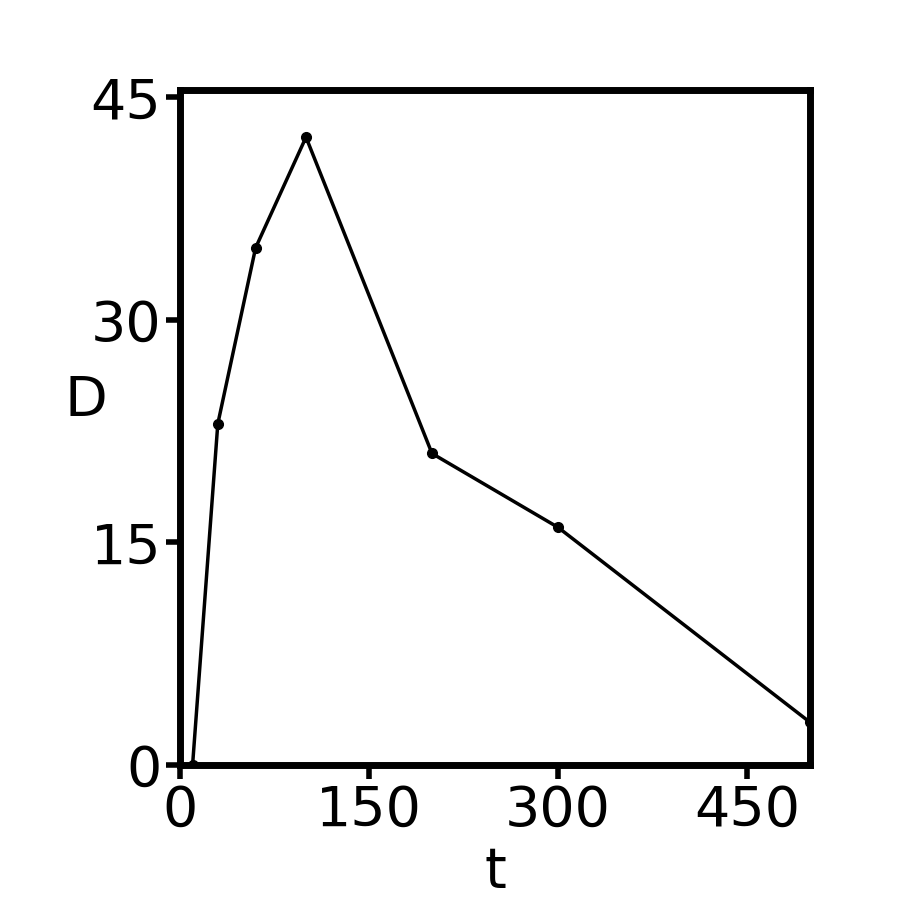}}
\caption{ (a) Comparison between the percentage $N$ of the Bohmian trajectories that have reached for the first time the area of the ergodic pattern as a function of time for the wavepacket around the unstable (blue) and the stable (red) triple periodic orbits (for $\epsilon=0.49$). (b) The  distance $D$ between the blue and red curves at the same times. }\label{fig7}
\end{figure}

The global manifestation of the present special case is shown in Fig.~\ref{fig7} (blue curve), where one can see that although all trajectories are chaotic, they approach their asymptotic ergodic pattern  more rapidly than in the stable period-3 case. In the same figure we show for comparison the corresponding times of the stable triple periodic orbit (red). {We see that the proportions associated with the unstable orbit are larger than those of the stable orbit at any given time.} We conclude that   the onset of Bohmian chaos is faster in the unstable periodic orbit case.

\begin{figure}[H]
\centering
\subcaptionbox{}{\includegraphics[scale=0.2]{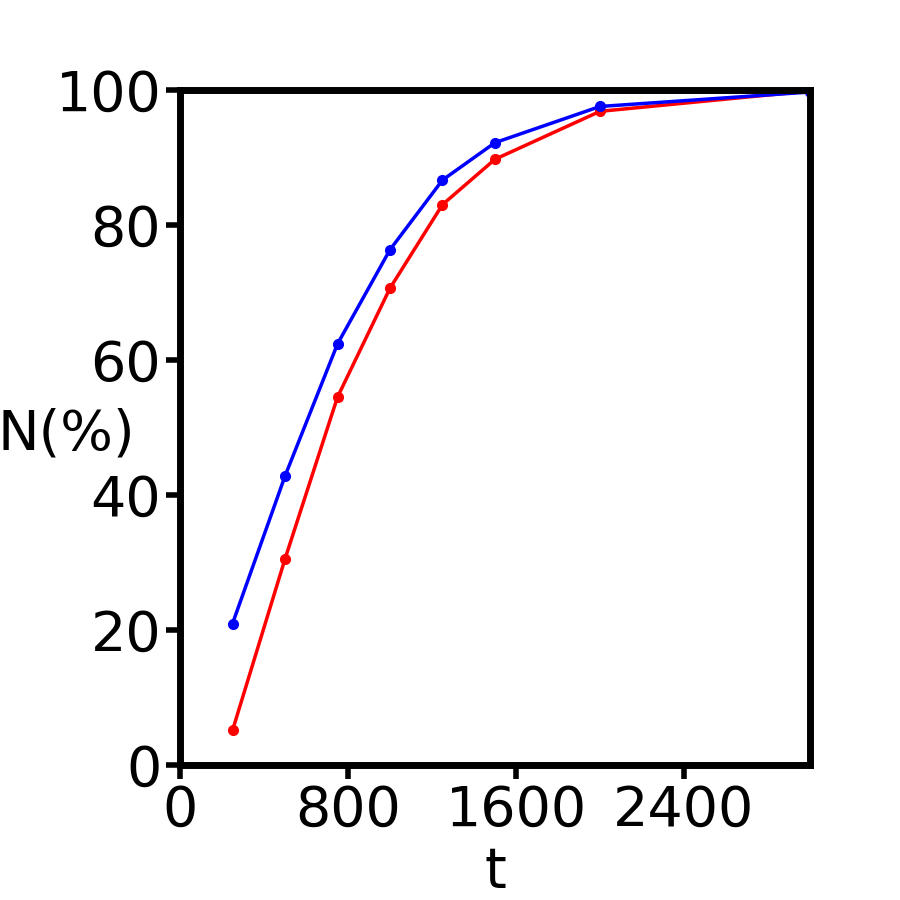}}
\subcaptionbox{}{\includegraphics[scale=0.2]{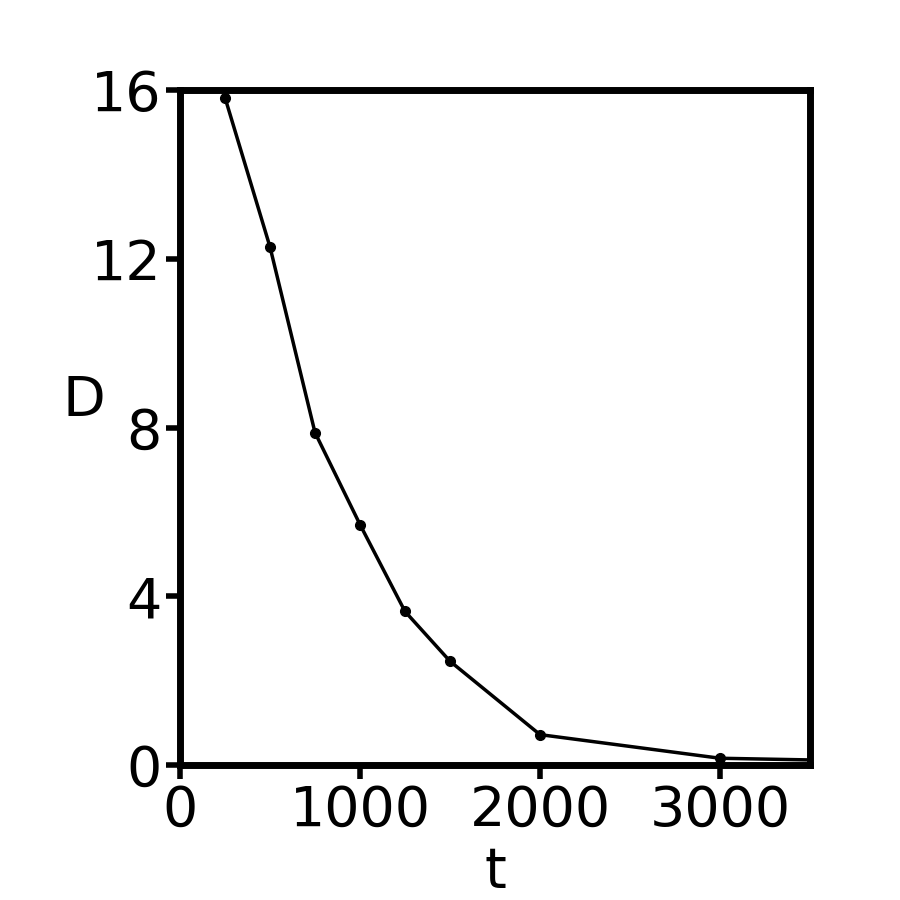}}
\caption{(a) As in Fig.~\ref{fig6} for $\epsilon=0.5017$ (stable period-3 orbit before the bifurcation) and $\epsilon=0.5025$ (unstable period-3 orbit after the bifurcation). The blue points have larger values at any given $t$. (b) The corresponding  difference between the red and the blue curves as a function of time. In this case $D$ goes smoothly to zero as time increases.} \label{fig8}
\end{figure}

To further examine the validity of our result regarding the difference in the times associated with stable and unstable orbits of the same period, it is instructive to investigate what happens when a stable periodic orbit becomes unstable at a bifurcation. In Fig.~\ref{fig8} we present our results for two wavepackets: one corresponding to the period-3 stable periodic orbit at $\epsilon=0.5017$, just before the bifurcation, and one corresponding to the period-3 unstable periodic orbit generated just after the bifurcation ($\epsilon=0.5025$). We again find that the stable orbit (blue curve) generally has larger times than the unstable one (red curve). In other words, the Bohmian trajectories are highly sensitive to the bifurcation and  immediately display a  change in their dynamics.

\begin{figure}[H]
\centering
\subcaptionbox{}{\includegraphics[scale=0.22]{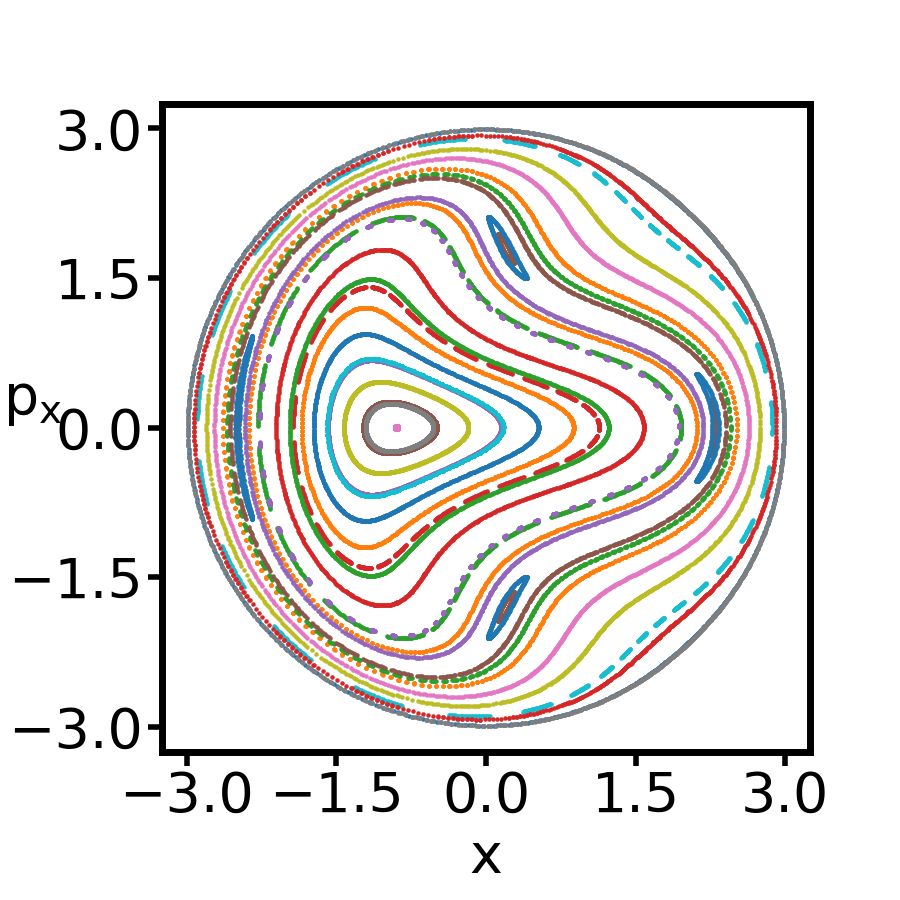}}
\subcaptionbox{}{\includegraphics[scale=0.22]{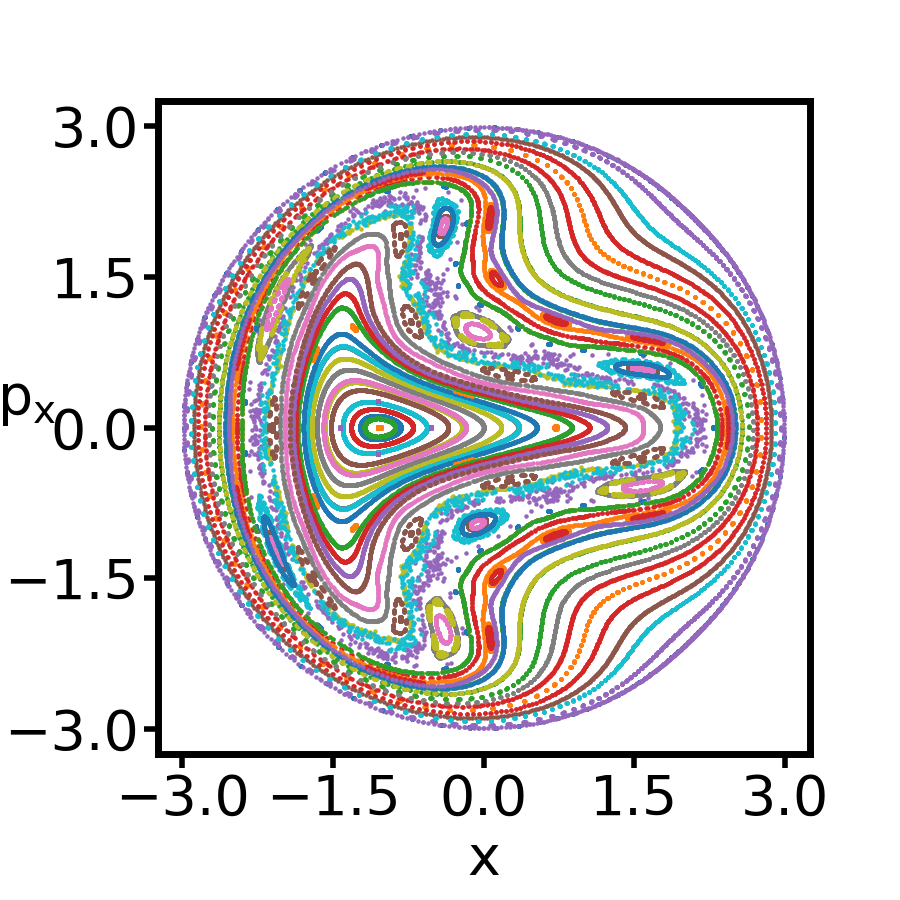}}
\subcaptionbox{}{\includegraphics[scale=0.22]{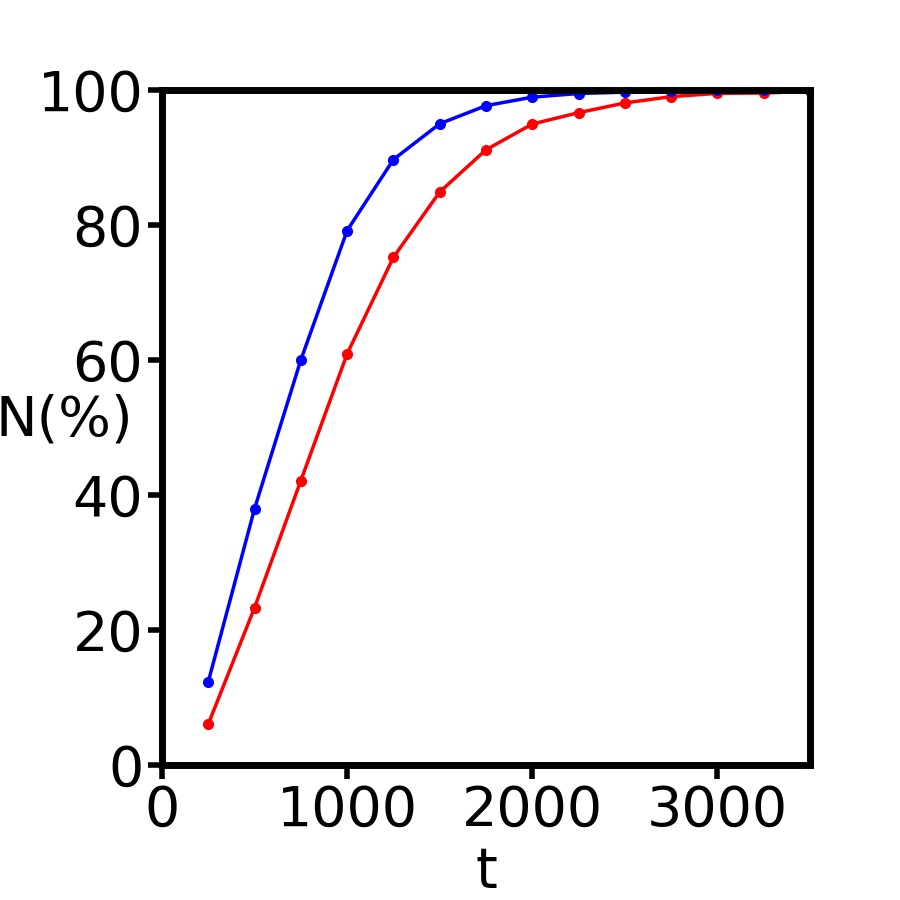}}
\subcaptionbox{}{\includegraphics[scale=0.22]{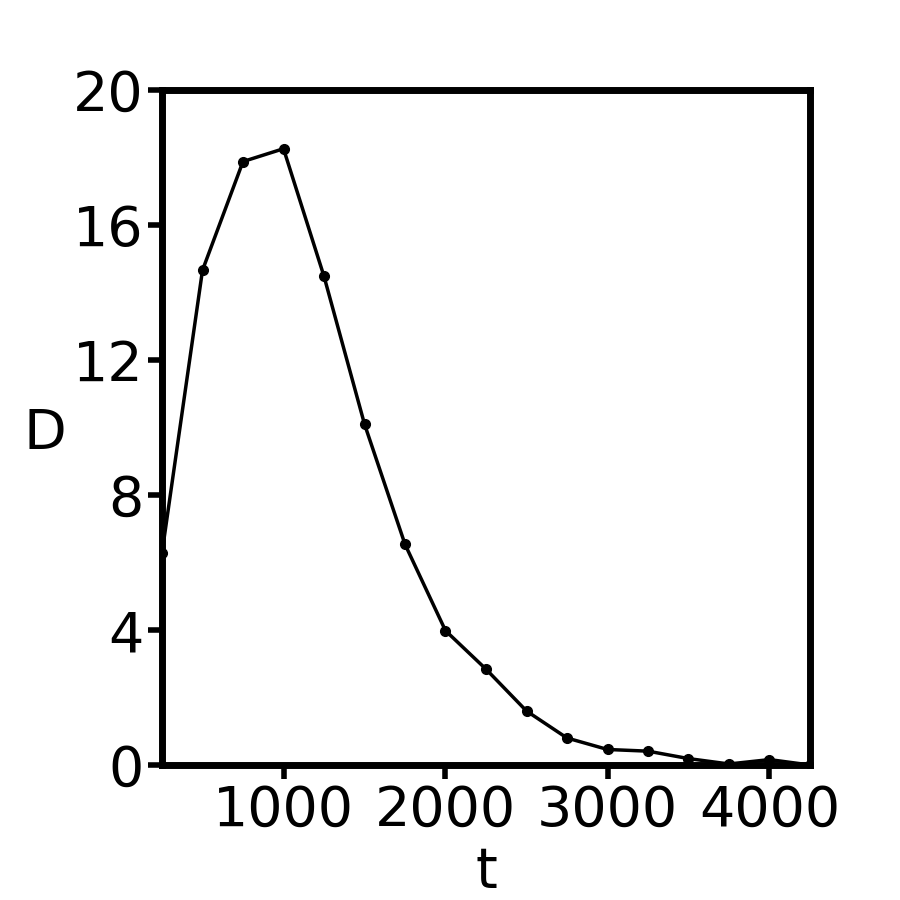}}
\caption{The Poincar\'{e} surface of section for $\omega_1=1,\omega_2=\sqrt{2}$ and $E=4.4967$ for (a) $\epsilon=0.29$ and (b) $\epsilon=0.32$.  In (a) we observe the period-4 stable periodic orbit (gray) which undergoes a pitchfork bifurcation to two symmetric stable period-4  orbits (cyan and brown) and two unstable periodic orbits between them in (b). (c) The corresponding times for one of the two stable periodic orbits of (b) and one of the unstable periodic orbits for $\epsilon=0.32$. Again the stable periodic orbit has larger times than the unstable one. (d) The absolute distance $D$ between the blue and red curves of (c).} \label{fig9}
\end{figure}

In order to check the generality of our results we also studied  a case with different frequencies. In particular, in  Figs.~\ref{fig9}(a) and (b) we show two characteristic Poincar\'{e} surfaces of section for $\omega_x=1$ and $\omega_y=\sqrt{2}$ and  two different values of the perturbation parameter, $\epsilon=0.29$ and $\epsilon=0.32$.   Between them a bifurcation of the period-4 periodic orbit takes place changing the period-4 stable periodic orbit into an unstable one and generating eight new stable periodic orbits with twice its period. Again we find that the stable periodic orbit (red) has larger times for generating Bohmian chaos than the unstable one (Figs.~\ref{fig9}(c),(d)).

\begin{figure}[H]
\centering

\subcaptionbox{}{\includegraphics[scale=0.20]{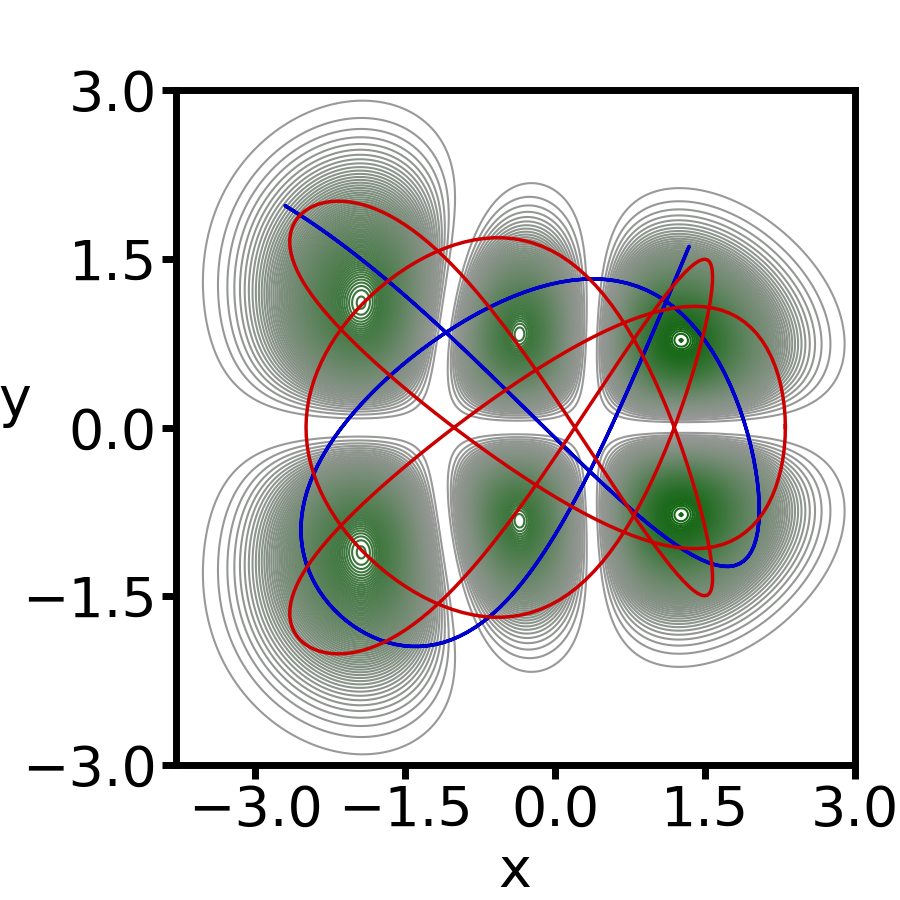}}\\
\subcaptionbox{}{\includegraphics[scale=0.11]{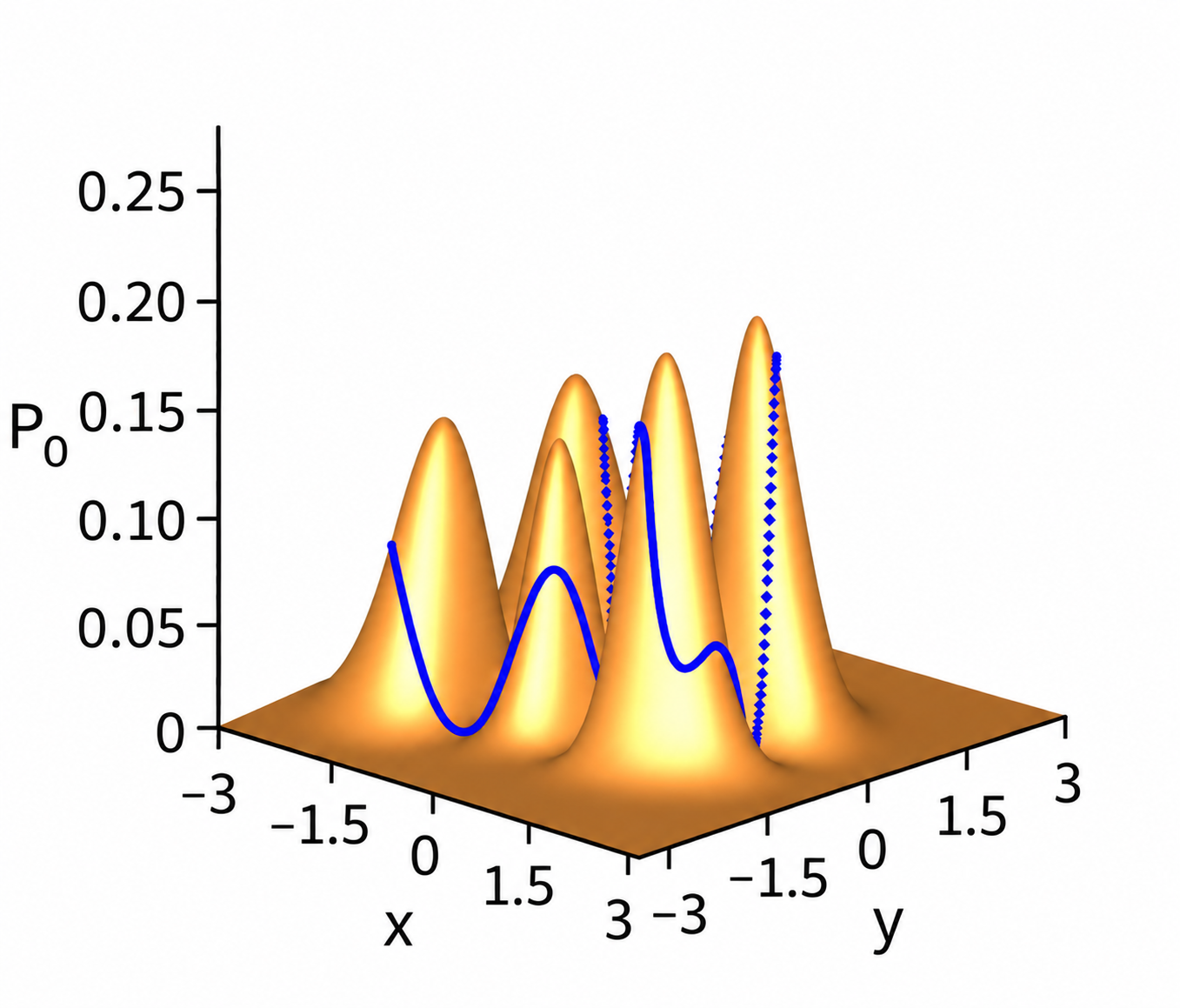}}
\subcaptionbox{}{\includegraphics[scale=0.11]{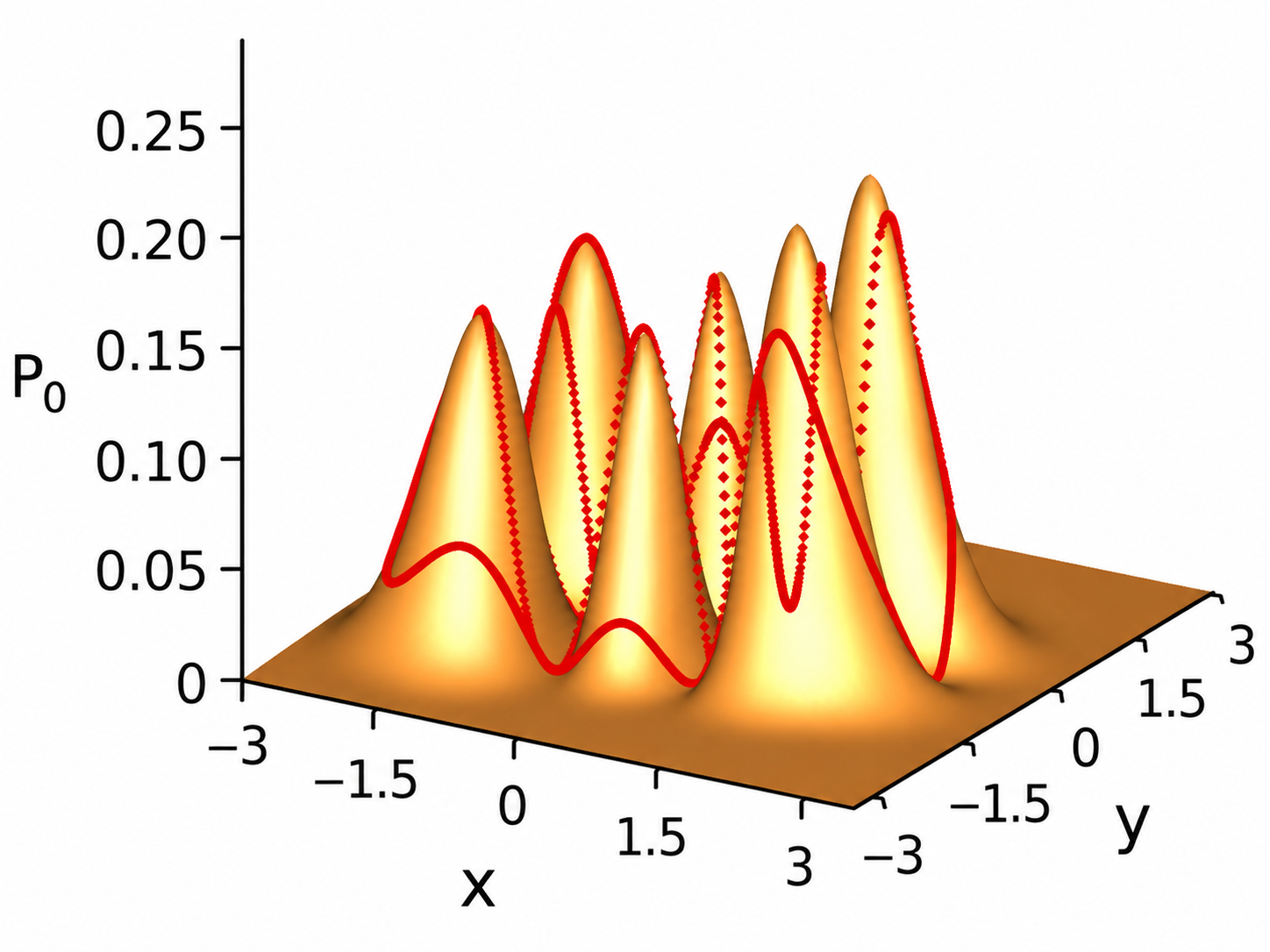}}
\caption{(a) Contour plot of the probability density of  the eigenstate of the perturbed system with the closest energy to that of the Poincar\'{e} surface of section ($E=4.488$) for $\epsilon=0.29$. Superimposed we see the stable (red) and the unstable (blue) periodic orbits of period 4. In (b) and (c) we show the corresponding 3D analogues of (a) for the unstable and stable orbits.} \label{fig10}
\end{figure}

In Fig.~\ref{fig10} we give the behaviour of an unstable (blue) and a stable (red) periodic orbit of period 4, as in Fig.~\ref{fig9}(a), for  $\epsilon=0.29$ for the perturbed eigenstate with the closest energy to that of the classical orbit.  We see that both the stable and unstable periodic orbits become significantly broader and more intricate than in the $\omega_{2}=\sqrt{3}$, $\epsilon=0.49$ case.

\subsection{Resonant cases}

In order to further complete our study and to test the generality of our results beyond the nonresonant frequency ratios examined in the previous sections, we now consider two representative resonant cases: the $1\!:\!1$ resonance with $\omega_x=\omega_y=1$ and the $2\!:\!1$ resonance with $\omega_x=2$ and $\omega_y=1$.

\begin{figure}[H]
\centering
\subcaptionbox{}{\includegraphics[scale=0.2]{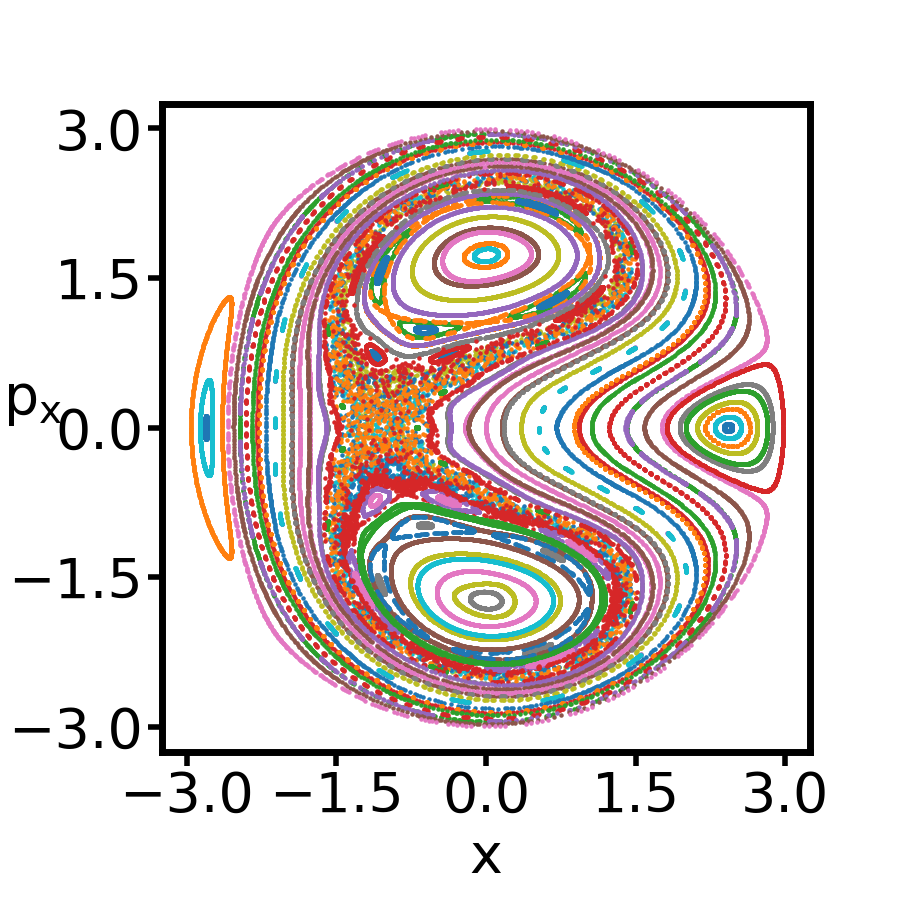}}\\
\subcaptionbox{}{\includegraphics[scale=0.2]{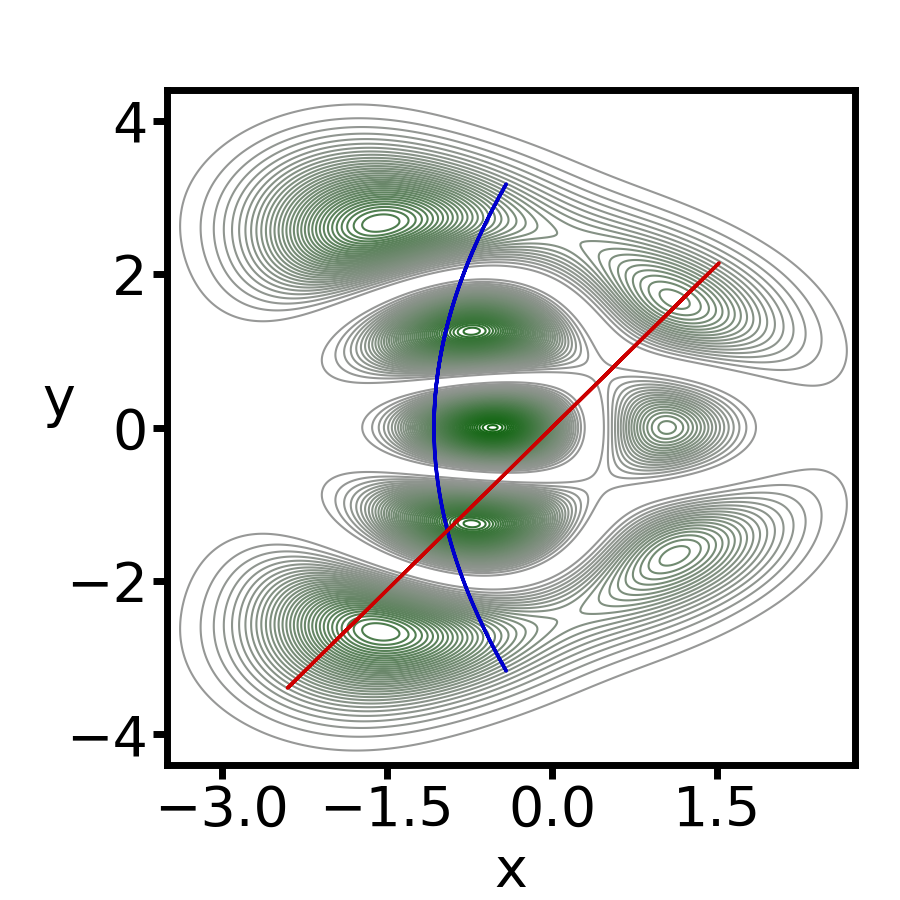}}
\subcaptionbox{}{\includegraphics[scale=0.2]{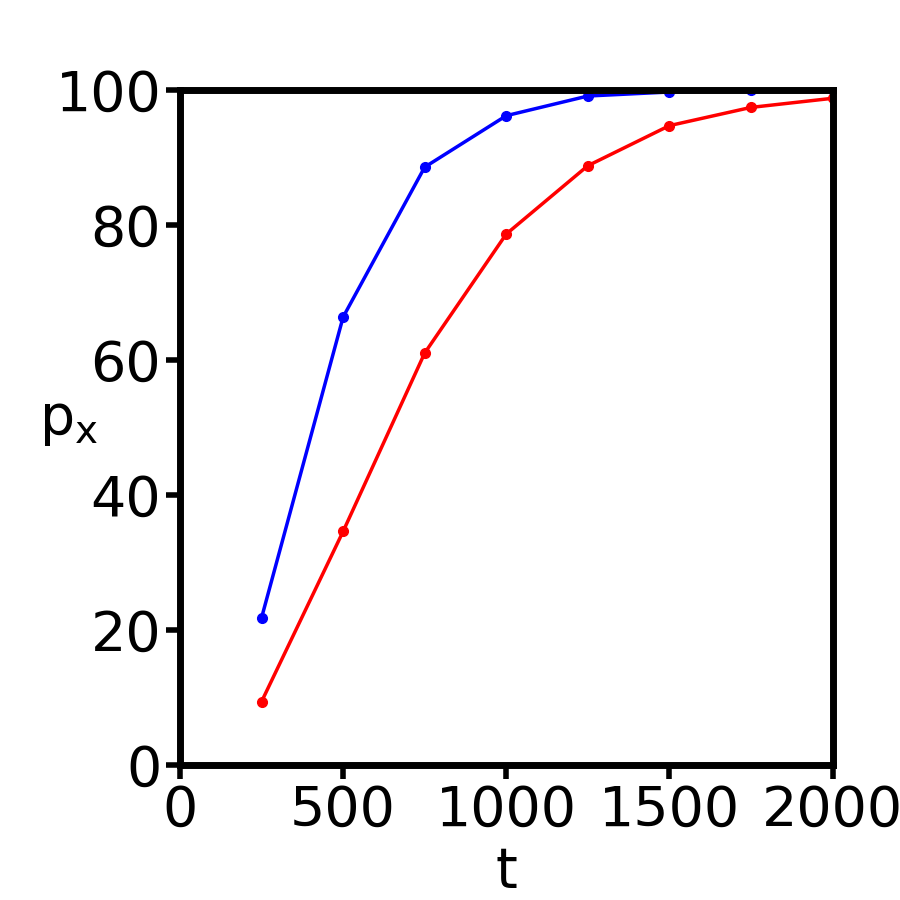}}
\caption{(a) Poincar\'{e} surface of section for the resonant case $\omega_x=\omega_y=1$ with $\epsilon=0.15$ and $E=4.4967$.
(b) Contour plot of the eigenstate of the corresponding quantum system with energy closest to the classical one ($E=4.692$), with the unstable (blue) and stable (red) periodic orbits superimposed (starting at $x=-1.07182,\ p_x=0$ and $x=0,\ p_x=1.73142$, respectively).
(c) Percentage $N(t)$ of Bohmian trajectories (Born ensemble of 5000 particles) that have reached for the first time the boundary of their long-term ergodic pattern as a function of time. The wavepacket centred around the unstable periodic orbit (blue curve) exhibits a systematically faster onset of Bohmian chaos than the one centred around the stable periodic orbit (red curve).} \label{fig11}
\end{figure}

We first consider the $1:1$ resonant case with perturbation strength $\epsilon=0.15$ and total energy $E=4.4967$.
The corresponding Poincar\'e surface of section (Fig.~\ref{fig11}(a)) exhibits a strongly mixed phase--space structure.
Three prominent period--1 periodic orbits are clearly identified: an unstable periodic orbit intersecting the surface of section at $(x=-1.07182,\ p_x=0)$ and two stable periodic orbits located at $(x=0,\ p_x=\pm 1.731417965300)$.
The contour plot of the quantum eigenstate with energy closest to the classical one ($E=4.692$) (Fig.~\ref{fig11}(b)) shows visible scarring associated with  an unstable and a stable periodic orbit, with the enhancement along the unstable orbit being more sharply localized, while the density around the stable orbit is broader.
The Bohmian dynamics  (Fig.~\ref{fig11}(c)) gives the proportion $N$ of orbits (out of a wavepacket of 5000 Bohmian particles  distributed according to Born's rule) that have reached the limits of their chaotic-ergodic pattern up to time t as a function of time. We see that the wavepacket centred around the unstable periodic orbit (blue curve) reaches its ergodic regime systematically faster than the wavepacket centred around the stable periodic orbit (red curve).

\begin{figure}[H]
\centering
\subcaptionbox{}{\includegraphics[scale=0.2]{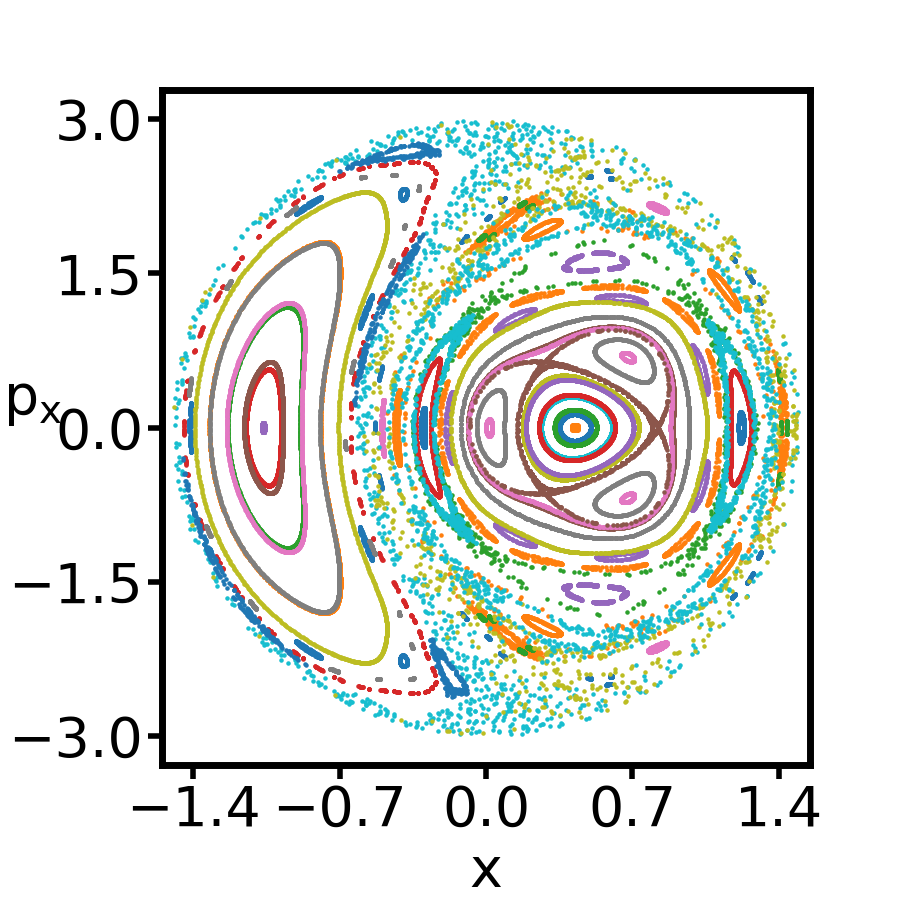}}\\
\subcaptionbox{}{\includegraphics[scale=0.2]{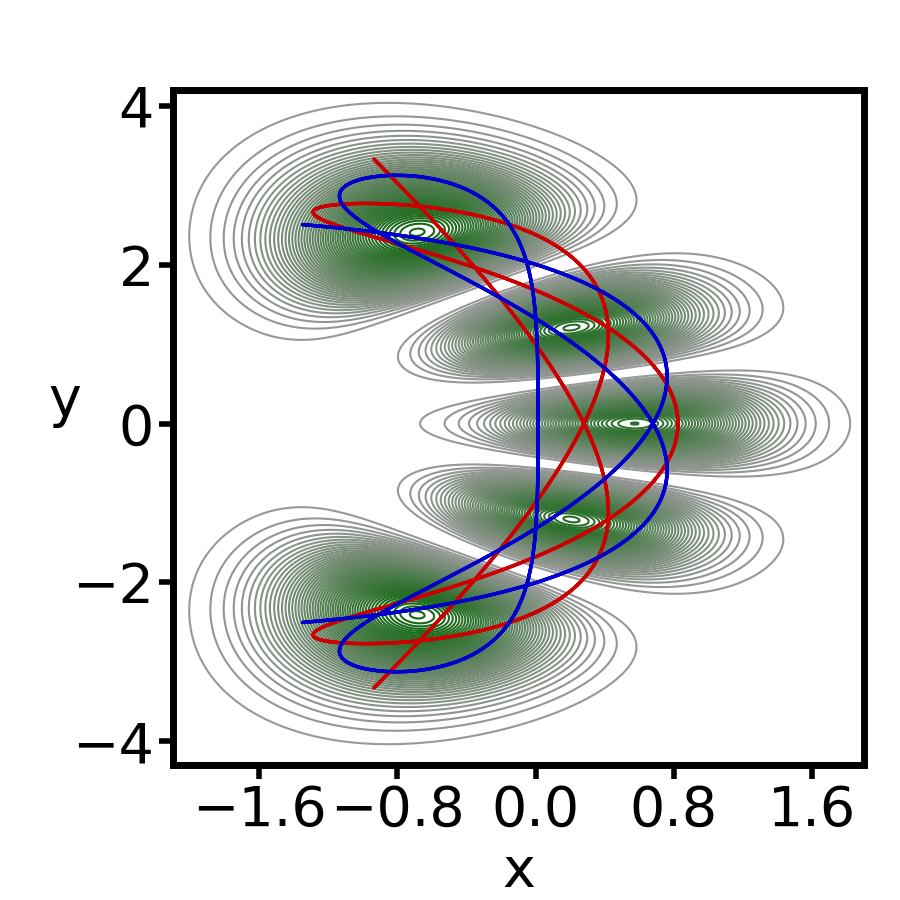}}
\subcaptionbox{}{\includegraphics[scale=0.2]{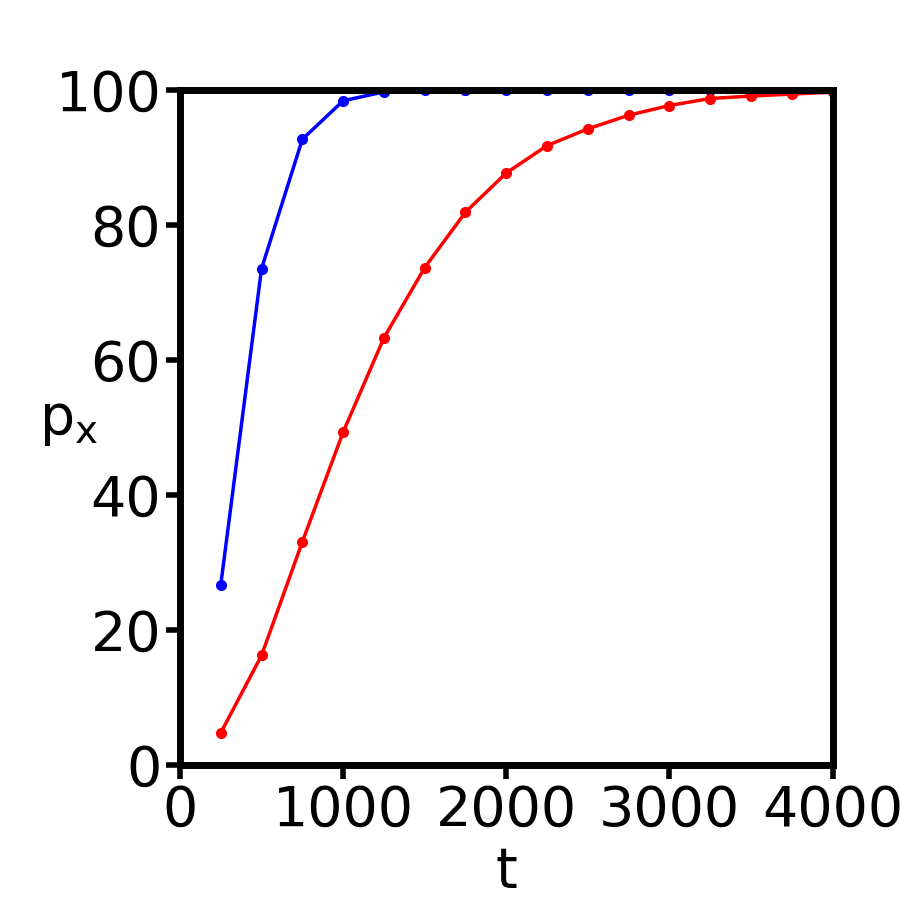}}
\caption{{(a) Poincar\'{e} surface of section for the 2:1 resonant case $\omega_x=2, \omega_y=1$ with $\epsilon=0.27$ and $E=4.4967$.
(b) Contour plot of the eigenstate of the corresponding quantum system with energy closest to the classical one ($E=4.224$), with the unstable (blue) and stable (red) periodic orbits superimposed (starting at $x=0.8230,\ p_x=0$ and $x=0.6789,\ p_x=0.6868$, respectively).
(c) Percentage $N(t)$ of Bohmian trajectories (Born ensemble of 5000 particles) that have reached for the first time the boundary of their long-term ergodic pattern as a function of time. Again in the wavepacket centred around the unstable periodic orbit (blue curve) we find a faster onset of Bohmian chaos than that centred around the stable periodic orbit (red curve).}} \label{fig12}
\end{figure}

We now turn to the $2:1$ resonant case with $\omega_x=2$ and $\omega_y=1$, for $\epsilon=0.27$ and the same total energy.
The Poincar\'e surface of section (Fig.~\ref{fig12}(a)) again displays a mixed phase--space structure with coexisting stable and unstable periodic orbits.
In this case we focus on a pair of period--4 periodic orbits: an unstable orbit intersecting the surface of section at $(x=0.823087738885,\ p_x=0)$ and a stable orbit located at $(x=0.678879230007,\ p_x=0.686799461917)$.
The  perturbed eigenstate $\tilde{\Phi}_5$ with $E=4.224$ (Fig.~\ref{fig12}(b)) exhibits pronounced scarring patterns aligned with both periodic orbits.
Nevertheless, the Bohmian dynamics (Fig.~\ref{fig12}(c)) follows the same qualitative behaviour, with the wavepacket centred around the unstable periodic orbit again displaying a systematically faster approach to ergodicity.

Both resonant cases confirm that the observed connection between classical stability, quantum scarring and Bohmian chaotic dynamics is robust both in resonant and nonresonant systems. For more technical details on the accuracy of the calculations see the Appendix B.

\section{High-energy regime and approach to the semiclassical limit}
{
The results presented above were obtained at relatively low energies, around $E\simeq4.5$. Although these calculations clearly reveal the different Bohmian dynamics associated with stable and unstable periodic orbits, it is important to examine whether the same behaviour persists at higher energies, where the de Broglie wavelength becomes smaller compared with the characteristic spatial scale of the classical motion. Such a comparison provides a more stringent test of the observed quantum-classical correspondence and allows us to assess whether the main conclusions remain valid closer to the semiclassical regime.}

A simple estimate of the relevant length scale can be obtained from the classical turning points. For the period-three orbit, which lies close to the $y=0$ axis, the turning points can be approximately estimated from the harmonic part of the Hamiltonian by setting $p_x=0$, giving $x_{\mathrm{tp}}\simeq\pm\sqrt{2E}$. The corresponding spatial extent is therefore roughly $L\simeq2\sqrt{2E}$. In the units used throughout this work, with $m=\hbar=1$, the characteristic momentum is of order $p\simeq\sqrt{2E}$, and the corresponding de Broglie wavelength is $\lambda_{\mathrm{dB}}\simeq2\pi/\sqrt{2E}$. At the energies considered in the main part of the paper, $E\simeq4.5$, this rough estimate gives $L\simeq6$ and $\lambda_{\mathrm{dB}}\simeq2.1$, corresponding to $\lambda_{\mathrm{dB}}/L\simeq0.35$. Thus, the wavelength is still a sizeable fraction of the characteristic classical length scale, indicating that this regime is not particularly close to the semiclassical limit.

{The above estimate is intended only as a rough characterization of the relevant scales. In the perturbed system, the actual periodic orbit need not follow the turning-point geometry of the unperturbed harmonic oscillator, and both its spatial extent and momentum distribution are determined by the full nonlinear dynamics. For this reason, we extended the calculations to the substantially higher energy $E=32$ and evaluated these quantities directly from the corresponding numerical period-three trajectory.}

\begin{figure}[H]
\centering
\subcaptionbox{}{\includegraphics[scale=0.2]{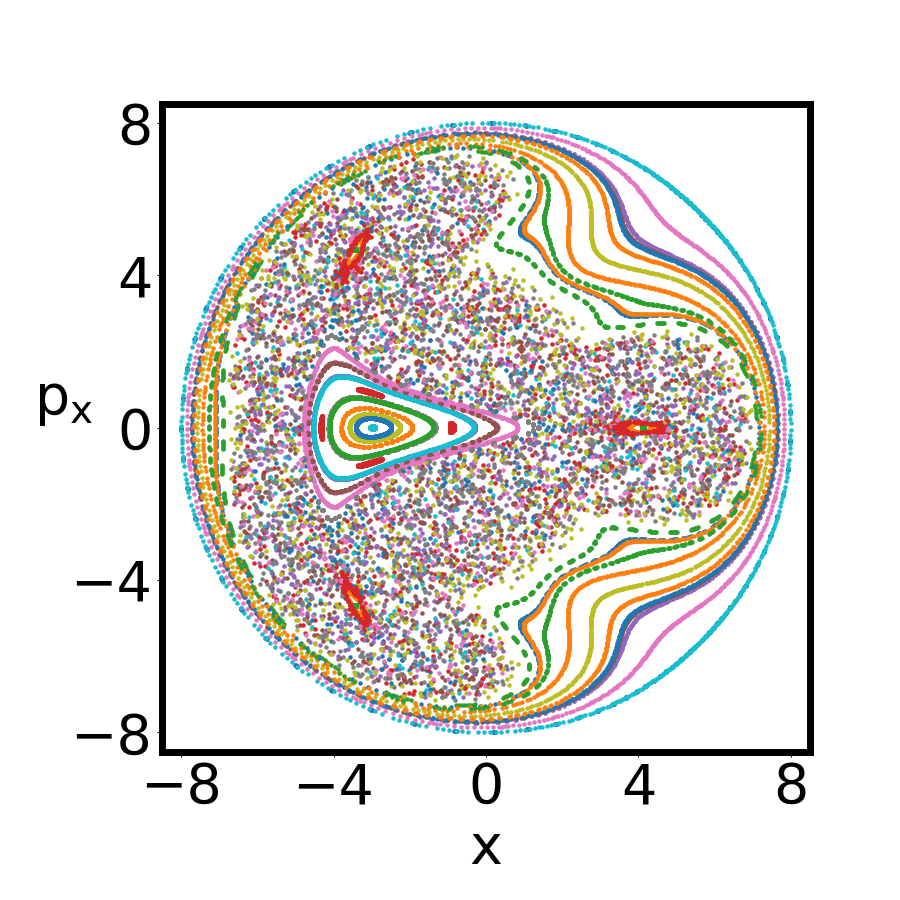}}
\subcaptionbox{}{\includegraphics[scale=0.2]{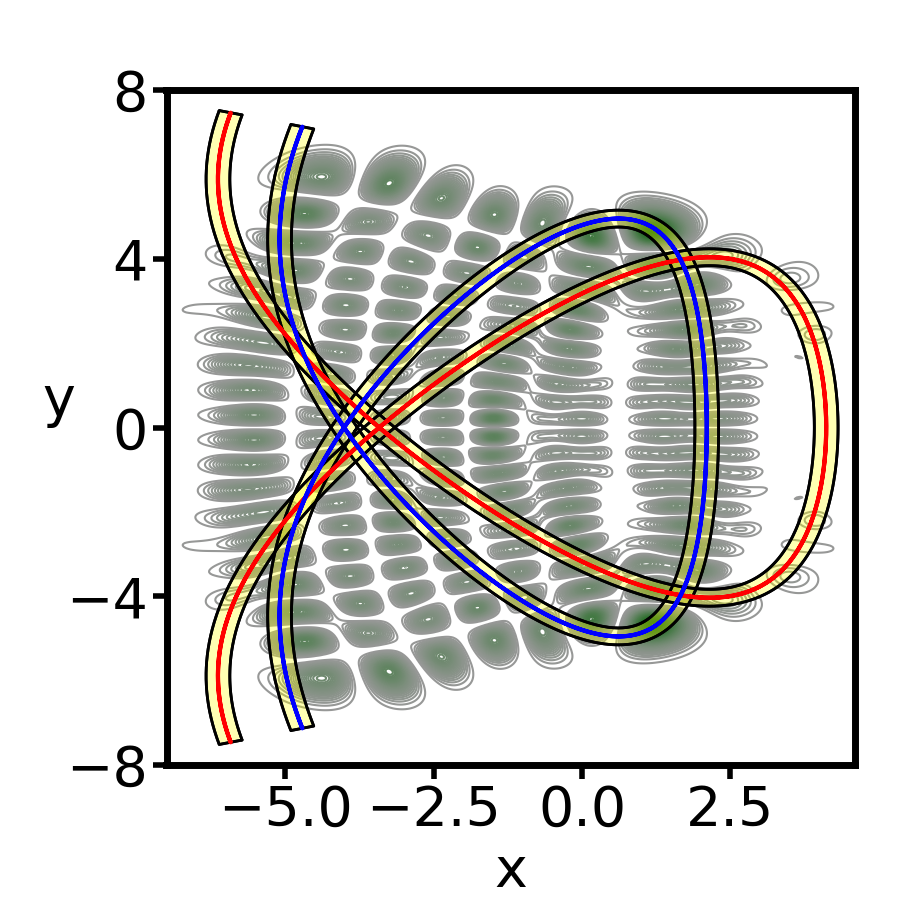}}
\subcaptionbox{}{\includegraphics[scale=0.2]{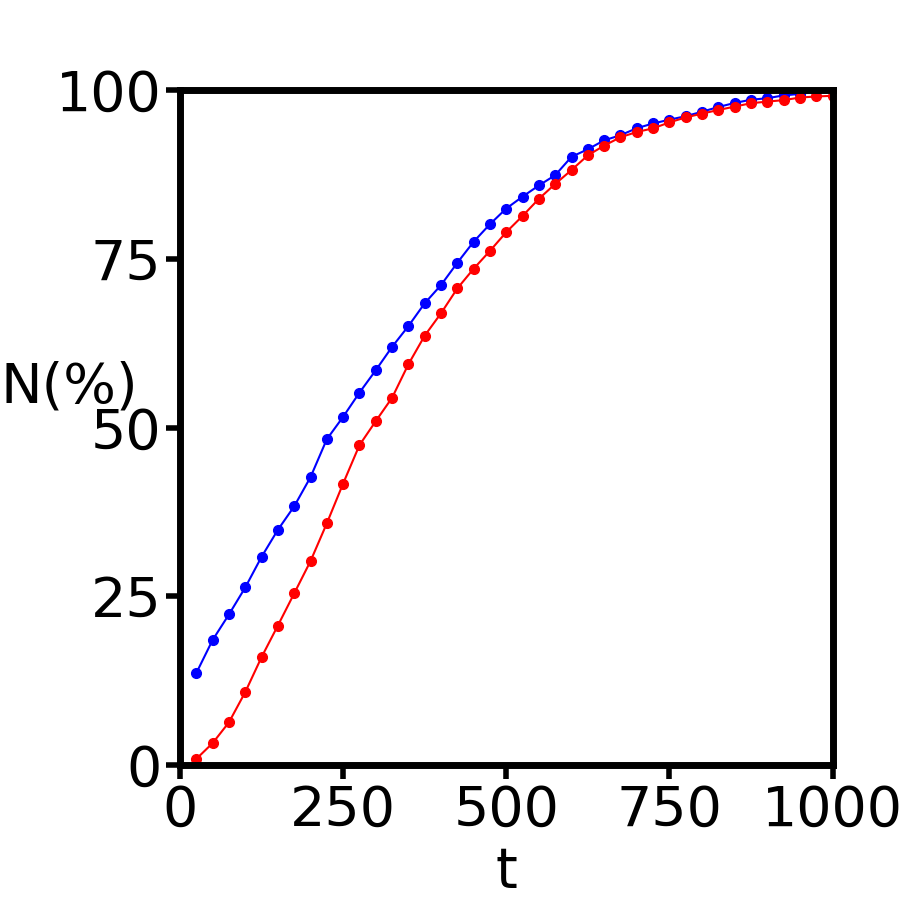}}
\caption{{(a) The Poincar\'e section of the classical system with $E=32$ ($\omega_x=1,\omega_y=\sqrt{2}, \epsilon=0.125$). We focus on the two period-3 periodic orbits whose form is  shown in (b) (unstable at $x = -4.00769, p_x = \pm 2.99837$ and $x =  2.1025776, p_x =  0$ and stable at $x = -3.4161, p_x = \pm 4.6641$ and $x =  4.1149, p_x = 0$). (b) Contour plot of the 394th eigenstate with eigenvalue $32.198$ which has the largest $P_{unst}/P_{st}$ inside the two tubes of width $w=0.4$ around the periodic orbits. (c) The percentage of the Bohmian trajectories of the wavepackets around the stable (red) and unstable (blue) periodic orbits that have explored for the first time the area of their limiting ergodic pattern as a function of time. } } \label{fig13a}
\end{figure}

{The three intersections of the period-three orbit of the Poincar\'e section for $E=32$ and $\epsilon=0.125$ (Fig.~\ref{fig13a}(a)) are found at
$x=-4.0077, p_x=\pm 2.9984$ and $x=2.10258, p_x=0$.
The full numerical trajectory extends over
$x_{\min}=-5.0952$, $x_{\max}=2.1026$ and
$y_{\min}=-7.1307$, $y_{\max}=7.1307$ (Fig.~\ref{fig13a}(b)).
To characterize its spatial extent without relying on the approximate turning-point construction above, we define}

\begin{equation}
L_{\mathrm{orbit}}=\max_{t,t'}\left|\mathbf{r}(t)-\mathbf{r}(t')\right|.
\end{equation}

{For the $E=32$ trajectory this gives $
L_{\mathrm{orbit}}=14.2615.$
We obtain the corresponding characteristic momentum directly from the numerical trajectory using
\begin{equation}
p_{\mathrm{char}}=
\sqrt{\left\langle p_x^2+p_y^2\right\rangle_t},
\end{equation}
which yields $p_{\mathrm{char}}=5.3591$. The corresponding de Broglie wavelength is therefore $\lambda_{\mathrm{dB}}=\frac{2\pi}{p_{\mathrm{char}}}=1.1724$.
Consequently, the ratio between the quantum wavelength and the characteristic spatial scale of the classical orbit is}
\begin{equation}
\frac{\lambda_{\mathrm{dB}}}{L_{\mathrm{orbit}}}=0.0822.
\end{equation}

This value is substantially smaller than the rough estimate of $\lambda_{\mathrm{dB}}/L\simeq0.35$ obtained at $E\simeq4.5$. Thus, the $E=32$ calculation corresponds to a regime in which the de Broglie wavelength is much smaller relative to the spatial extent of the classical periodic orbit, providing a  more robust test of the persistence of the observed Bohmian dynamics closer to the semiclassical limit.

The increase in energy comes at a considerable computational cost. The original calculations used a basis truncated at $n_x+n_y\leq10$, corresponding to 66 unperturbed basis states. For the $E=32$ calculations, the truncation is increased to $n_x+n_y\leq41$, resulting in 903 basis states. Consequently, the Hamiltonian matrix becomes more than an order of magnitude larger in each dimension, while the total number of matrix elements increases by almost two orders of magnitude. The increased basis size also makes the subsequent construction and time evolution of the wavefunction, as well as the calculation of the Bohmian velocity field, substantially more demanding.

Figure~\ref{fig13a}(b) shows the contour plot of the probability density $\rho=|\psi|^2$ of the  eigenstate with the most pronounced scarring of the unstable orbit. Although the probability density is visibly concentrated in the vicinity of both the stable and unstable orbits, it is not possible to determine by visual inspection which of the two orbits exhibits the stronger localization. We therefore introduce a tubular neighbourhood around each periodic orbit in order to quantify the probability associated with it.

{For a classical periodic orbit $\mathbf{r}(t)$, the tube is constructed by displacing each point of the orbit along the direction perpendicular to its local velocity,}
\begin{equation}
\mathbf{R}(t,n)=\mathbf{r}(t)+n\mathbf{e}_{\perp}(t),
\qquad -w\leq n\leq w .
\end{equation}

{The total probability contained within the tube surrounding the $k$th eigenstate is then calculated as}
\begin{equation}
P_{\mathrm{tube}}^{(k)}
\int_0^T\int_{-w}^{w}
|\psi_k(\mathbf{R}(t,n))|^2
J(t,n)dndt ,
\end{equation}
{where $J(t,n)$ is the corresponding Jacobian of the transformation from $(t,n)$ to the two-dimensional configuration space. The same procedure is applied separately to the stable and unstable periodic orbits. Their relative localization is quantified by the ratio}

\begin{equation}
R_k=
\frac{P_{\mathrm{unst}}^{(k)}}
{P_{\mathrm{st}}^{(k)}} .
\end{equation}

By calculating the ratio $R_k$ for various eigenstates around the energy 32 we found that it becomes maximum and equal to $2.038$ for $k=394$, where the corresponding eigenvalue is $E= 32.198$.

Then, by counting again the percentage of trajectories, inside two 5000-particle realizations of Born's rule in initial wavepackets around the triple stable and unstable periodic orbits, that explore for the first time the area of the final ergodic pattern we find again that in the case of the unstable orbit we have a quicker onset of chaos (Fig.~\ref{fig13a}(c)). Namely, for each given time the percentage $N$ is greater than that of the stable orbit. 

Thus this high-energy calculation further supports our previous findings, showing that Bohmian trajectories associated with unstable periodic orbits explore their long-time ergodic patterns faster than those associated with stable periodic orbits.

\section{Conclusions}

The scarring phenomenon in quantum-mechanical wavefunctions has been extensively studied in the literature. Most previous works have focused on scarring associated with unstable classical periodic orbits. However, similar scars can also emerge along classically stable periodic orbits.

In this work we explored in detail how classical periodic orbits of a perturbed 2D oscillator influence the Bohmian dynamics of the corresponding quantum system. Our analysis combined classical phase space structures, quantum scarring phenomena, and large ensembles of Bohmian trajectories in order to clarify how the stability properties of periodic orbits influence quantum scarring and quantum chaos in the Bohmian framework.

The classical system exhibits the characteristic features of a nearly integrable Hamiltonian with a nonlinear perturbation: large stability islands, chaotic regions, and an intricate network of stable and unstable periodic orbits. Among these, the period--1 and period-3 orbits play a prominent role, particularly in the regime where the stable period-3 orbit undergoes a bifurcation and gives rise to an unstable period-3 orbit and two stable period-6 orbits. This rich classical structure provides the natural background against which quantum scarring and Bohmian chaos must be understood.

On the quantum side, the diagonalization of the Hamiltonian reveals a small set of perturbed eigenstates near the energy of a classical period--3 orbit. In the first model ($\omega_x=1,\omega_y=\sqrt{3}$) two of these states exhibit clear and well--localized scarring aligned with the unstable periodic orbit (while the third state shows no such enhancement). The scarring associated with the stable period-3 orbit is also present but it is much more diffuse, filling the interior of the classical stability islands rather than concentrating along the orbit itself. These observations indicate a general principle, namely that unstable orbits imprint sharper, more coherent scars, while stable orbits generate broader probability accumulations.

This difference in the spatial localization of the scars provides a possible mechanism for the different Bohmian dynamics observed in the two cases. Around unstable periodic orbits, the quantum probability density is concentrated more sharply along the orbit, whereas around stable periodic orbits the enhancement is distributed more broadly over the surrounding stability island. Consequently, the Bohmian trajectories associated with the unstable orbit can leave the vicinity of the periodic structure and explore the accessible region more rapidly, while those associated with the stable orbit remain correlated with the surrounding stability-island structure for longer times.

Bohmian dynamics provides a natural framework for relating quantum scarring patterns to the time-dependent motion of Bohmian particles. In this work, we employ minimum-uncertainty wavepackets which, on the one hand, closely approach classical behavior by being spatially localized around the corresponding classical periodic orbits, and, on the other hand, are constructed as superpositions of a large number of energy eigenstates of the perturbed system. As a result, these wavepackets naturally fulfill the necessary conditions for the emergence of Bohmian chaos, which relies on the interference of multiple eigenstates.

We found that all Bohmian trajectories ultimately become chaotic and explore an ergodic domain determined by the support of the wavefunction. In most cases this domain is connected.

A central result of our study is the systematic comparison of the times required for the Bohmian trajectories to reach the boundaries of their long--term ergodic region for the first time (ergodic times). For wavepackets centred around unstable periodic orbits, these times are consistently shorter than those associated with stable periodic orbits of the same period. In other words, the Bohmian flow associated with an unstable orbit becomes chaotic faster.

The sensitivity of these times becomes especially clear when examining bifurcations. As soon as a stable period--3 orbit becomes unstable, the Bohmian ergodic times associated with its wavepacket drop sharply.

This behaviour persists in a second model with frequency ratio ($\omega_x=1,\omega_y=\sqrt{2}$) and in two resonant cases ($\omega_x=\omega_y=1$ and $\omega_x=2,\omega_y=1$) where a period--4 orbit undergoes a bifurcation. Again we find that the stable periodic orbit has larger times than the unstable one.

These additional calculations provide a test of the generality of the observed behaviour beyond the original non-resonant model. In the $\omega_x=1,\omega_y=\sqrt{2}$ case, a different period--4 bifurcation leads from a stable to an unstable periodic orbit and generates new stable orbits. The same qualitative behaviour is also obtained in the resonant $1:1$ and $2:1$ cases. Thus, the faster Bohmian spreading associated with unstable periodic orbits is observed for different frequency ratios and for different bifurcation scenarios.

The robustness of this connection is further supported by the higher-energy calculation at $E=32$. In this case the ratio between the de Broglie wavelength and the characteristic spatial extent of the classical periodic orbit is $\lambda_{\mathrm{dB}}/L_{\mathrm{orbit}}=0.0822$, substantially smaller than the corresponding estimate at $E\simeq4.5$. The persistence of the observed behaviour in this regime provides an additional indication that the relation between classical stability, quantum scarring and Bohmian dynamics is not restricted to the relatively low-energy regime considered in the main calculations.

In the Appendix A we cover the case of energy-centred wavepackets. These packets, although distributed over nearly the same eigenstates as their geometrically centred counterparts, exhibit markedly slower chaotic spreading. The reason lies in their spectral composition: namely the probability densities have most of their weight on a small group of eigenstates with very small energy gaps, producing slowly varying relative phases in the wavefunction and therefore a slowly evolving Bohmian velocity field. As a result, the trajectories remain localized near the periodic orbits for significantly longer times before reaching the ergodic domain. This mechanism operates for both stable and unstable orbits.

Finally, in Appendix B we provide more details for the accuracy of the calculation of the energy spectrum of our perturbed system and its dependence on the ratio of the frequencies $\omega_x,\omega_y$.

Taken together, our results indicate a structural connection between classical periodic-orbit stability, the spatial localization of quantum scars, and the time scale of Bohmian chaotic dynamics. Unstable periodic orbits are associated with sharper scarring and faster exploration of the corresponding ergodic domain, whereas stable periodic orbits produce broader scarring and longer ergodic times. The agreement between the different non-resonant and resonant models and the higher-energy calculation suggests that this connection is robust and deserves further investigation in other nonlinear Hamiltonian systems and for other types of bifurcations.

\section*{Appendix A}

In the main text we worked with Gaussian wavepackets whose centroids were placed at the geometric position of classical periodic orbits.

However, a spatially centred Gaussian does not in general have the same mean energy as the corresponding classical orbit. For wavepackets of the form \eqref{wp}, the average energy is
\begin{align}
\nonumber
\langle E\rangle &= \int\!\!\int \Psi_{coh}^\dagger \hat{H}\Psi_{coh}\,dx\,dy \\ &
= H_0 + \frac{1}{4}\left(
\omega_1^2\sigma_x^2
+ (2\epsilon x_0 + \omega_2^2)\sigma_y^2
+ \sigma_x^2
+ \sigma_y^2
\right),
\end{align}
and therefore depends on the centroid coordinates $(x_0,y_0,p_{x0},p_{y0})$. Therefore there is a difference between the desired classical energy and the actual mean energy of the packet.

To address this, we also consider a second type of initial condition, where the parameters of the Gaussian  are adjusted so that the mean energy matches that of the periodic orbit. We refer to these as energy-centred wavepackets and repeat the Bohmian simulations using Born-rule ensembles of 5000 trajectories.

\begin{figure}[H]
\centering
\subcaptionbox{}{\includegraphics[scale=0.2]{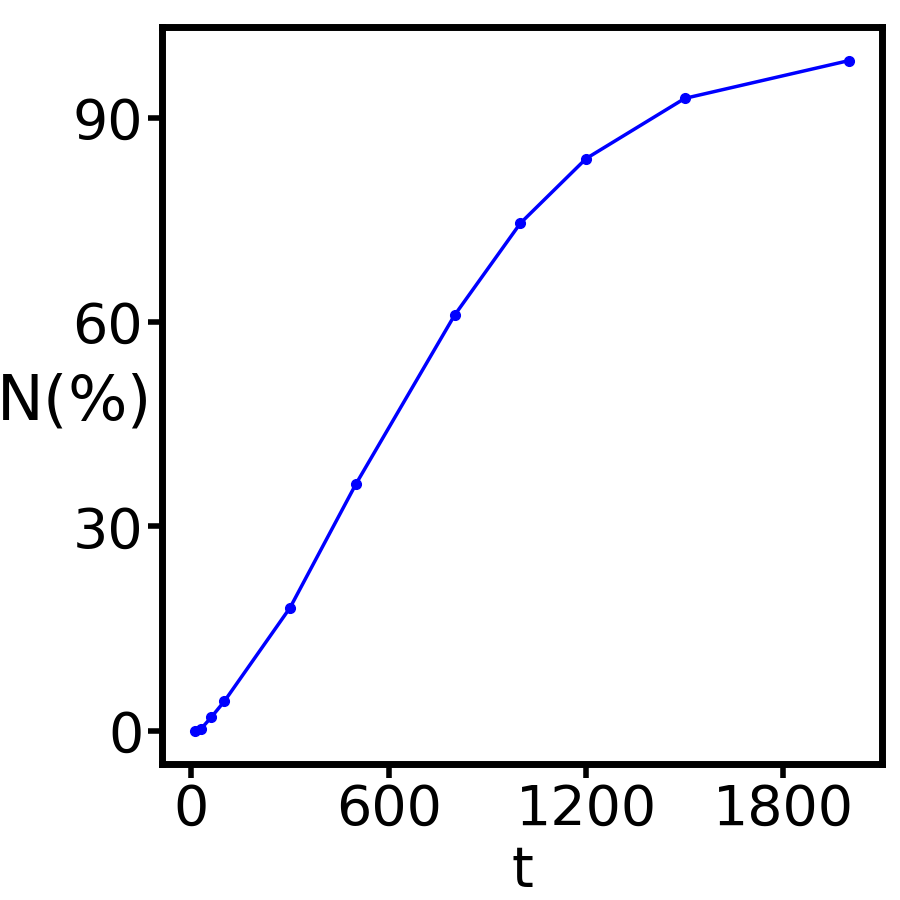}}
\subcaptionbox{}{\includegraphics[scale=0.2]{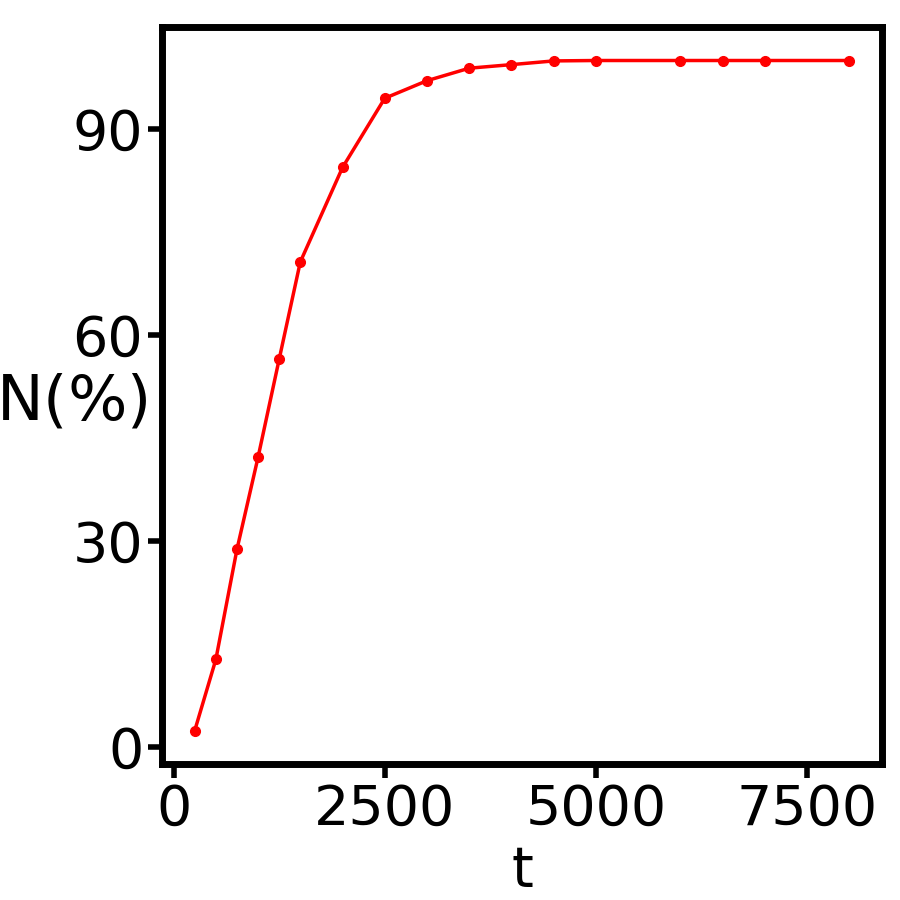}}
\caption{The percentage $N$ of the Bohmian trajectories that have reached for the  first time the area of the ergodic pattern as a function of time for the energy centred packet (a) of the unstable triple periodic orbit and (b) the stable triple periodic orbit. The corresponding times are much larger than those of Fig.~\ref{fig5}(b) and Fig.~\ref{fig5}(d).}\label{fig13}
\end{figure}

The results for both the unstable and the stable period-3 periodic orbits are shown in Fig.~\ref{fig13}. It is evident that, for energy-centred wavepackets, the chaotic Bohmian trajectories require significantly longer times to reach the boundaries of their eventual ergodic domain for the first time, compared with the trajectories arising from geometrically-centred wavepackets.

\begin{figure}[H]
\centering
\subcaptionbox{}{\includegraphics[scale=0.25]{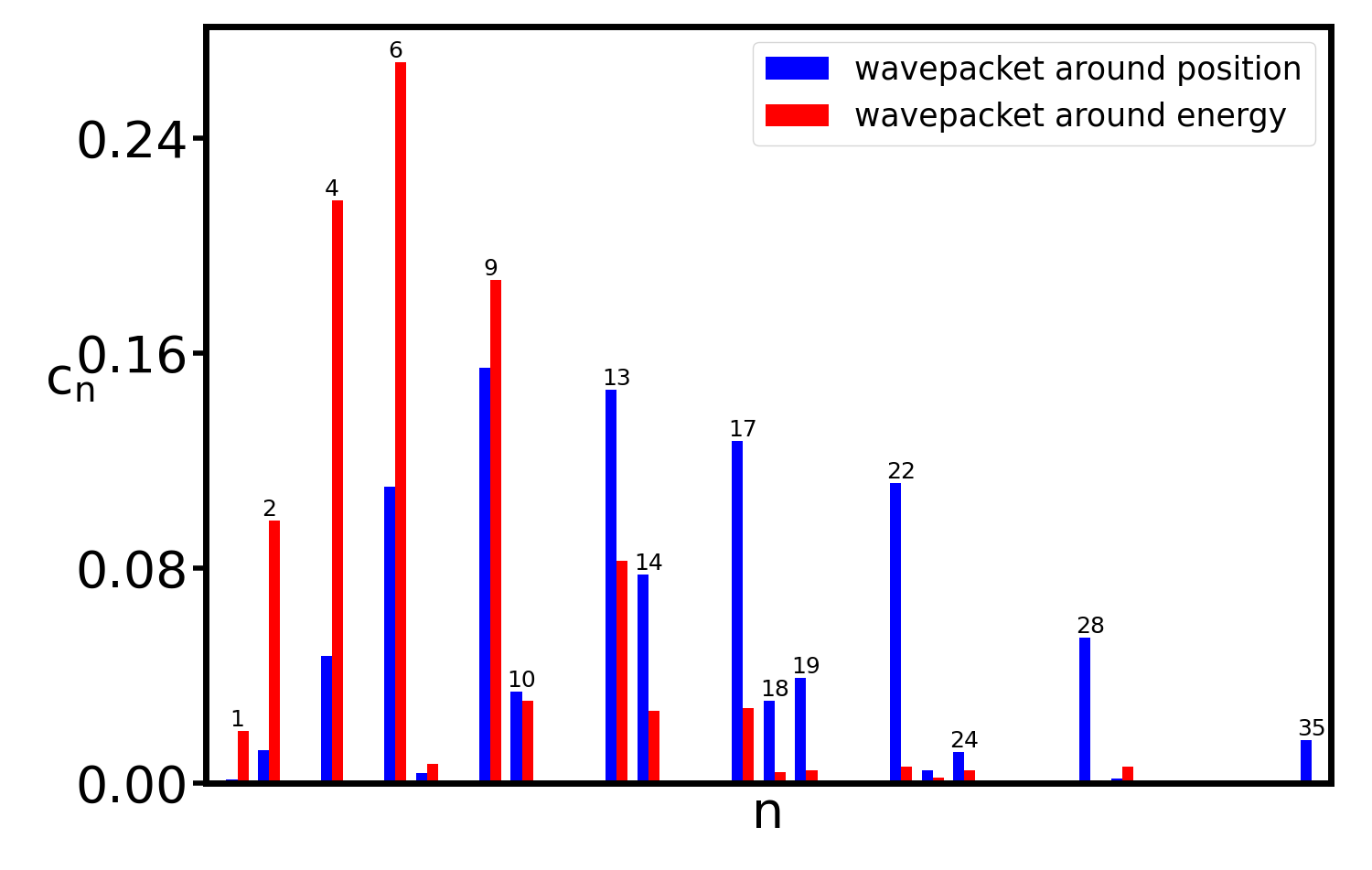}}
\subcaptionbox{}{\includegraphics[scale=0.25]{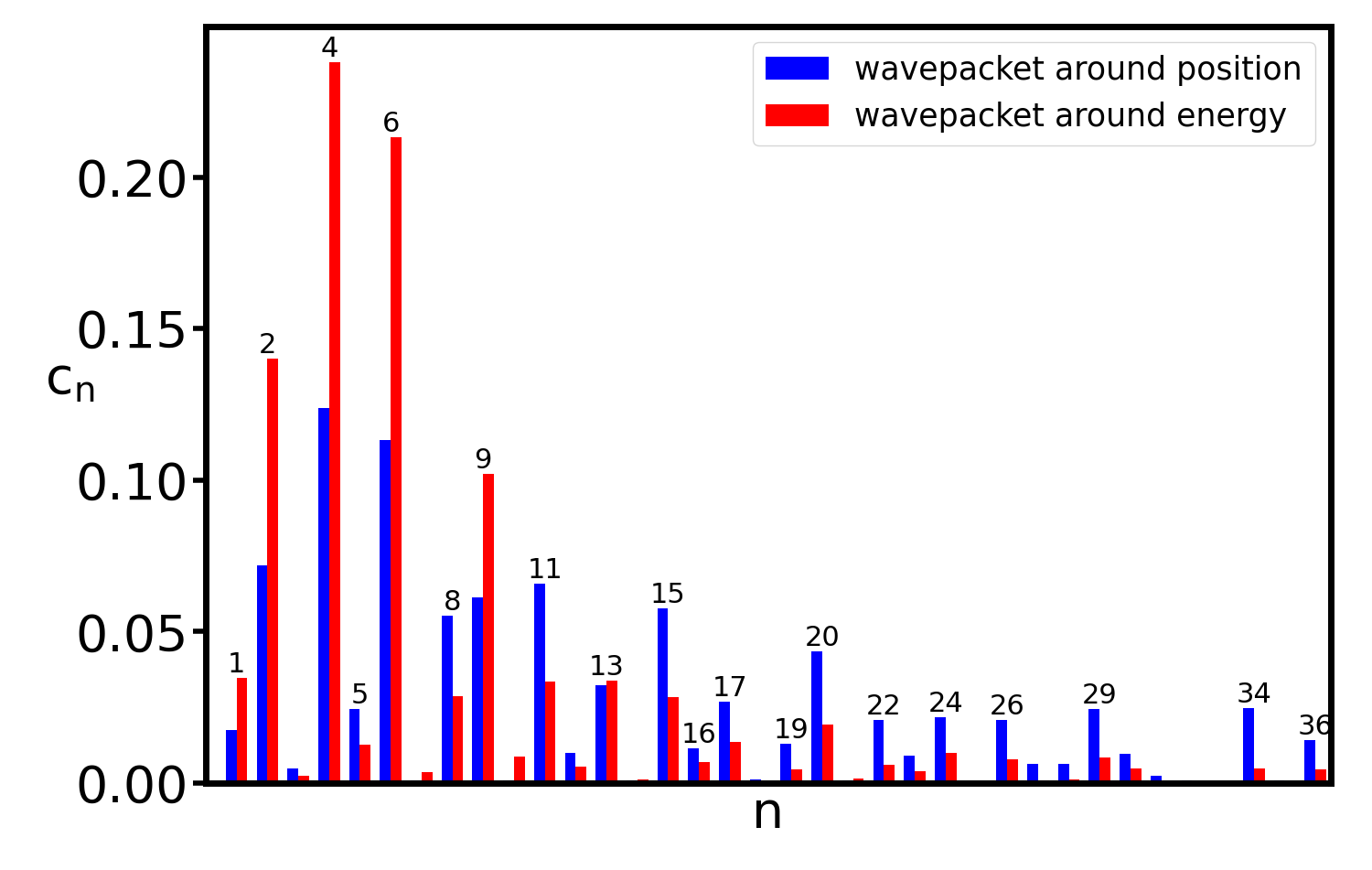}}
\caption{(a) Statistical weights of the perturbed eigenstates for wavepackets centred at  the geometric position of the triple unstable periodic orbit (blue) and at its classical energy (red). (b) The same comparison for the triple stable periodic orbit.}\label{fig14}
\end{figure}

To understand this difference, we examine the projections of the two packets onto the perturbed eigenstates $\{\tilde{\Phi}_n\}$. Figure~\ref{fig14}(a) shows that both packets involve approximately the same set of eigenstates, but with different statistical weights. The position-centred packet spreads its amplitude across several scarred eigenstates (e.g.\ $n=6,9,13,17,22$), producing a more localized initial density near the periodic orbit. The energy-centred packet instead places most of its weight on a smaller group of nearby eigenstates, with the scarred state $\tilde{\Phi}_6$ being dominant.

Since the Bohmian velocity field depends on the time evolution of the relative phases, the relevant quantities are the energy differences $E_n - E_m$. The dominant eigenstates of the energy-centred packet have small energy gaps, which lead to slowly varying phase differences and therefore a slowly changing velocity field. As a result, the trajectories remain near the periodic orbit for a longer time before entering the final ergodic regime.

Similarly, in Fig.~\ref{fig14}(b) we have the leading states in the two types 
of wavepackets for the stable triple periodic orbit. In both panels the
energy centred wavepacket shows a clearer concentration of weight on a small set
of dominant eigenstates compared to the position centred one. However, the lower
panel (corresponding to the triple stable periodic orbit) involves a
substantially larger number of perturbed eigenstates overall. This broader
mixing is consistent with the structure of the eigenfunctions associated with
stable periodic orbits, which are typically less sharply localized than the
scarred states near unstable orbits. As a result, any wavepacket constructed
around the stable orbit necessarily draws amplitude from a wider portion of the
spectrum. The qualitative difference between position  and energy centred
packets remains the same, but it appears on top of this larger background of
mixed eigenstates.

In summary, the longer ergodic times of the energy-centred packet arise from its broader initial spatial support and from the smaller energy gaps among the eigenstates that carry most of its weight.

\section*{Appendix B}

We note here that in the resonant cases a larger truncation of the Hilbert space was required compared to the nonresonant ones.

The unperturbed energy levels of the two--dimensional anisotropic harmonic oscillator are given by
$E_{n_x,n_y}=\omega_x(n_x+1/2)+\omega_y(n_y+1/2)$.
For irrational frequency ratios, such as $\omega_x=1$, $\omega_y=\sqrt{3}$ or $\omega_x=1$, $\omega_y=\sqrt{2}$, the typical energy separation
$|E_{n_x,n_y}-E_{n_x',n_y'}|$ between unperturbed basis states remains of order unity.
In this case, the perturbation $\epsilon x y^2$ mainly couples states with relatively large energy differences, so that the corresponding mixing amplitudes,
which scale as
$\langle n_x,n_y|\epsilon x y^2|n_x',n_y'\rangle / (E_{n_x,n_y}-E_{n_x',n_y'})$,
remain small.
As a result, the effective mixing is limited and convergence of both eigenvalues and eigenfunctions is achieved already for moderate truncation sizes ($N=10$ for $\omega_y=\sqrt{3}$ and $N=12$ for $\omega_y=\sqrt{2}$).

In contrast, for low--order resonant cases such as the $1\!:\!1$ and $2\!:\!1$ frequency ratios, the unperturbed spectrum contains sets of degenerate states which get weakly detuned in the presence of the perturbation. Namely,  the energy denominators
$|E_{n_x,n_y}-E_{n_x',n_y'}|$
become small for some sets of basis states giving rise to much larger mixing amplitudes. Consequently the corresponding perturbed eigenstates require a larger Hilbert space in order to be accurately represented ($N=14$ for $\omega_x=\omega_y=1$ and $N=13$ for $\omega_x=2,\omega_y=1$).

\section*{Acknowledgements}
This work was funded by the Sectoral Development Program ($O\Pi\Sigma 5223471$) of the Ministry
of Education, Religious Affairs and Sports, through the National Development Program (NDP) 2021-25.
It was conducted as part of project $200/1026$, supported by the Research Committee of the Academy
of Athens.

\bibliographystyle{elsarticle-num}
\bibliography{bibliography}

\end{document}